\documentclass{article}

\usepackage[T1]{fontenc}
\usepackage{authblk}
\usepackage[utf8]{inputenc}
\usepackage{newtxtext,newtxmath}
\usepackage{graphicx}
\usepackage[]{units}
\usepackage{enumitem}
\usepackage{booktabs}
\usepackage{url}
\usepackage{mathtools}
\usepackage[caption=false]{subfig}
\usepackage[numbers,square,sort]{natbib}
\usepackage[toc,page]{appendix}
\usepackage{hyperref}
\providecommand{\keywords}[1]{\textbf{\textit{Keywords: }} #1}

\begin{document}

\title{PolicySpace2: modeling markets and endogenous public policies}
\author[1,2]{Bernardo Alves Furtado}
\affil[1]{Institute for Applied Economic Research} 
\affil[2]{ National Council for Scientific and Technological Development}

\maketitle

\begin{abstract}
Policymakers decide on alternative policies facing restricted budgets and uncertain future. Designing public policies is further difficult due to the need to decide on priorities and handle effects across policies. Housing policies, specifically, involve heterogeneous characteristics of properties themselves and the intricacy of housing markets and the spatial context of cities. We propose PolicySpace2 (PS2) as an adapted and extended version of the open source PolicySpace agent-based model. PS2 is a computer simulation that relies on empirically detailed spatial data to model real estate, along with labor, credit, and goods and services markets. Interaction among workers, firms, a bank, households and municipalities follow the literature benchmarks to integrate economic, spatial and transport scholarship. PS2 is applied to a comparison among three competing public policies aimed at reducing inequality and alleviating poverty: (a) house acquisition by the government and distribution to lower income households, (b) rental vouchers, and (c) monetary aid. Within the model context, the monetary aid, that is, smaller amounts of help for a larger number of households, makes the economy perform better in terms of production, consumption, reduction of inequality, and maintenance of financial duties. PS2 as such is also a framework that may be further adapted to a number of related research questions.

\keywords{Public policies \and Real estate market \and Agent-based modeling \and Simulation \and Spatial analysis \and Metropolitan regions}
\end{abstract}

\section{Introduction}

\label{intro}
Policymakers main question could be posed as: "given a fixed amount of investment, how to prioritize among alternative policies in order to provide the most benefits for its population?" This task is specially difficult if it involves policies across different sectors and inputs that generate results in time and space, and includes economic uncertainties. 

Real estate markets is one that faces intricate complexity. It suffers influence from (a) economic cycles, interest levels and liquidity \citep{davis_nieuwerburgh_2015,leamer_housing_2015}, (b) households intertemporal decisions and changes \citep{glaeser_extrapolative_2017}, (c) local and foreign investors' interest \citep{saiz_brief_2019}, (d) land-use, zoning regulation, and permits \citep{glaeser_urban_2006}, (e) construction dynamics, and (f) location, location, location: such as job market spatial proximity, amenities and neighborhoods \citep{bourassa_spatial_2007,rosenthal_change_2015}. Besides this vast amount of interconnections, the dwelling as a marketable good itself pertains some singular characteristics \citep{whitehead_urban_1999,lambiri_real_2014}. They are durable and expensive, usually indivisible and with elevated transaction costs; mostly heterogeneous, and with monopolistic relative location. Finally, housing contributes to GDP (construction, renting, services) and constitutes a large part of the stock of wealth \citep{ibbotson_world_1985,morandi_novas_2016}.\footnote{\citet{morandi_novas_2016} estimates construction represented 1.8 of Brazilian GDP in 2014, and a total net fixed capital stock of nearly 8,000,000 (Million R\$ of 2010).} 

There is a large and consistent literature that handles most of these intricacies of the market. From urban economics and macroeconomics specifically, a vast array of studies follow the framework of \citet{dipasquale_markets_1992} \citep{hendershott_asymmetric_2010,steiner_estimating_2010}. More recently though, in the aftermath of the subprime housing crisis it has become clear that traditional models are insufficient to handle empirical volatility and excessive price variance \citep{davis_nieuwerburgh_2015,glaeser_housing_2015,shiller_subprime_2008,glaeser_extrapolative_2017}. 

These observed inefficiencies of typical modeling fostered alternative lines of investigation based on more empirical approaches \citep{davis_nieuwerburgh_2015,leamer_housing_2015,saiz_brief_2019} and with an emphasis on computational simulations \citep{dawid_agent-based_2014,dawid_agent-based_2018}. These models aim at replicating patterns and processes so that they might be useful tools in understanding mechanisms and evaluating \textit{ex-ante} policies. Real estate market models specifically have focused on macroprudential analysis in order to inform monetary authorities on how to prevent or reduce excessive volatility in the housing market \citep{baptista_macroprudential_2016,geanakoplos_getting_2012}. Policy-makers indeed face a daunting task of providing housing market stability at the national scenario as well as of ensuring that citizens have adequate, serviced, and affordable housing at the local level.  

Given this context, we propose PolicySpace2 (PS2) as a primarily endogenous computational agent-based model (ABM) that includes mortgage loans, housing construction, taxes collection and investments, with firms and households interacting in real estate, goods and services, and labor markets. PS2 is applied to 46 metropolitan regions in Brazil and serves the purpose of comparing local policies' investments over three alternative mechanisms to lower-income households: (a) housing acquisition and distribution , (b) rental payments over two years, and (c) a transfer of monetary aid. 

PS2 adds to the literature as it contains elements of three general approaches to modeling: (a) macroeconomics ABM -- summarized by \citet{dawid_agent-based_2018} into seven "families" of models which focus on country-level economics, with at most a regional approach; (b) land-use change \citep{filatova_agent-based_2009,parker_multi-agent_2003} -- which describes urban processes of expansion, land conversion from rural to urban and its different uses, such as commercial, residential, industrial ans service; and (c) transport and urban planning \citep{waddell_integrated_2018,horni_multi-agent_2016,zhuge_agent-based_2016}, which emphasize the interplay between transport expansions and urban land-use change, from a planning perspective. 

Furthermore, the contributions of PS2 simultaneously include a number of modeling characteristics. The model: (a) is open source, (b) uses official data at the intraurban level for the 46 major metropolitan regions of Brazil, (c) applies explicit spatial rules for three different markets, (d) includes a tax system at the municipal level, (e) is based on firms and households decision-modeling, and (f) whose policy experiment is implemented from endogenous demand and supply processes. 

The fact that PS2 is an agent-based model supports the observation of other dynamics resulting from policy interventions, besides the specific tests made in this paper. Future work might benefit from observing disaggregated results from: (a) households composition and location, (b) firms productivity and location, (c) migration, new household formation and demographics, (d) credit and financial liquidity, (e) labor market and selection processes that simultaneously consider qualification, distance, and access to public or private transport, and (f) the dynamics of construction and the real estate market. 

Validation and verification of the model comes from a successive and cumulative number of steps. Processes and rules are based on literature of previous models. Rules are tested in a structural sensitivity analysis \citep{goldstein_rethinking_2017}. Parameters have been tested exhaustively and perform robustly through a great variation of scenarios coming from different metropolitan regions' settings. Results are presented as an average of simulation runs. Further, 66 indicators help follow different aspects of the simulation.

In the midst of all of the processes involved in the simulation, we were able to calibrate PS2 to provide reasonable macroeconomic indicators: GDP, inflation, unemployment, and Gini coefficient that remain within expected values. Furthermore, even without the inclusion of any data referring to the real estate market itself, such as property characteristics, we were able to replicate the first half of the prices distribution for the case of Brasília effectively, for which we had empirical data to compare.

Running PS2 evidentiates the fact that the model captures the relevance of the housing market to the economy as a whole. Increase in households' savings, an influx of households, elevated productivity or a higher participation of households in the market; all increase GDP and quality of life within the model. 

Besides this introduction, the following section presents the modeling approaches to real estate market analysis and agent-based modeling and refers back to how PS2 contributes to existing literature. Methods describes the model and the policy experiment design. We then present the sensitivity analysis and validation, the results of the simulation and policy tests. We conclude with some final considerations. 

\section{Approaches to real estate market and modeling}\label{lit}

\subsection{Real estate markets}
\label{real}

Housing is a major component of capital stock \citep{causa_housing_2019}. The residence is also households' most expensive purchase \citep{dipasquale_urban_1996}, which may involve mortgage payments for a number of years. Affordability of financial costs or rental burden is a larger problem for developing countries, such as Brazil \citep{fundacao_joao_pinheiro_estatistica_2018}. It may be a socially desirable goal to prevent volatility, sudden cycles and disruptive rental markets \citep{ozel_macroeconomic_2019,nijskens_hot_2009}. 

Moreover, “[t]he housing market is a dynamic system of intricately woven interdependent processes” \citep[p.511]{jordan_agent-based_2012}. In addition to the specificities of the characteristics of the dwelling itself and the financial market as such, other mechanisms make real estate market analysis complicated: (a) neighborhood's perception, the buzz and in its impact on valuation \citep{galster_william_1996,jacobs_economy_1970}, (b) the ever-changing spatial urban context and scale in which the dwelling sit \citep{brueckner_why_1999,wheaton_commuting_2004,bettencourt_origins_2013}, and (c) the long-run dynamics of these continuous altering landscape, with rigid, slow-adjusting stock \citep{dipasquale_markets_1992,arnott_economic_1987}. 

Location relative to other residences \citep{fujita_spatial_1999,mills_advances_1987} as well as to the transport \citep{waddell_urbansim:_2002} system are also relevant to price dynamics. On top of it all, individual activities' time allocation and mobility management also play a role in households' decision-making towards housing \citep{arentze_incorporating_2010,moeckel_constraints_2017,zhuge_agent-based_2016}. 

Analysis of spatial patterns of firms and houses started in the 1960s with \citet{alonso_location_1964} and was synthesized by the Muth-Mills model \citep{brueckner_structure_1987}. Thereafter, the Central Business District (CBD) assumption was relaxed with the introduction of models with multiple equilibria \citep{fujita_multiple_1982}, increasing returns \citep{fujita_spatial_1999} and mixed landuse \citep{wheaton_commuting_2004}. \citet{NBERw29078} propose a "unified theory of cities" that encompasses increase returns and costly commuting. The authors make a sensitivity analysis over the parameter space to find that multiple equilibria is a common feature. They also suggest that increasing returns and lower commuting costs might either generate agglomeration or disagglomeration.

The definition of sub-markets considers spatially clustered dwellings that are close substitutes for each other whereas mostly distinct relative to dwellings in other sub-markets \citep{wheaton_does_2005,bourassa_defining_1999}. Urban economists have controlled for sub-markets including neighborhoods, and neighbourhood perceived values into their hedonic price functions \citep{bourassa_spatial_2007,furtado_neighbourhoods_2011,rosen_hedonic_1974}. 

Most housing policy programs reflect the idea that owning your house should be a household goal \citep{davies_right_2013}. As a consequence, public policy tends to enforce the idea that homeownership is better than renting \citep{causa_housing_2019}. In fact, \citet{causa_housing_2019} suggest that house ownership might help equalize the distribution of wealth, mainly because it is the most important household's asset. The authors also show that cross-country comparisons have found distinct levels of homeownership.

The calculus and decision-making between acquisition or renting is rather difficult, mainly because of the unknowns and uncertainties about future parameters, such as inflation or interest rates \citep{malmendier_2017}. Even under perfect certainty, best choice between house acquisition and renting depends on the externalities provided by renting vis-à-vis the variation in gain and losses associated with housing \citep{henderson_model_1983}. Despite the financial calculus, life cycles, i.e., endogenous household changes, also influence tenure choices \citep{andersen_motives_2011}. 

Furthermore, \citet{malmendier_2017} suggest that higher proportions of ownership tend to promote neighborhood engagement and social capital, leading to higher house prices. Renting, however, gives households more mobility and safety from housing price volatility. \citet{mcafee_machine_2017} go as far as to say that a new consensus around renting as the best option is forming. 

\subsection{Real estate and agent-based modeling}\label{real estate lit.}

Traditional modeling \citep{dipasquale_urban_1996} suggests a spatial equilibrium in which all of these conditions clear: (a) supply and demand in the purchasing and rental markets, (b) appreciation of the estate's value equivalent to premium of interests of the economy, and (c) salaries and amenities balance across other localities. More recent urban economics frameworks \citep{NBERw29078} still focus on general equilibrium, providing comparative statics analysis on a linear city. As much as the exercise sheds lights on the problem at hand, we argue that Agent-based Modeling (ABM) might act as a complementary tool to support policy counterfactual insights and enable prospective empirical comparisons. 

\citet{glaeser_extrapolative_2017} pinpoint that real estate empirical data do not follow all theoretical constraints needed by traditional modeling. The authors demonstrate that observed data contain relevant momentum, mean reversion and "excessive variance relative to fundamentals" \citep[p.1]{glaeser_extrapolative_2017}. Thus, even using historic data to make price forecasting is a rather difficult task that depends heavily on the amount and precision of available information. All in all, the literature suggests that real estate markets are inherently complex, encompassing the financial market, future expectations, intrinsic features of the property itself, location, utility to households and investors, and altering spatial context \citep{saiz_brief_2019, davis_nieuwerburgh_2015,leamer_housing_2015}. At the same time, theoretical tools seem to be insufficient to maneuver all of these elements together. 

Agent-based model (ABM) refers to the construction of computational models in which agents follow explicit rules, and interact with other agents and the environment. One of the first applications to economics was the El Farol problem proposed by \citet{arthur_inductive_1994} to discuss bounded rationality. By early 2000s, a consensus was consolidated around its meaning and usability \citep{tesfatsion_agent-based_2006,lebaron_agent-based_2006}. Dawid and Gatti described benchmarks and best practices in what they called 'families' of macroeconomic agent-based models in the end of the 2010s \citep{dawid_agent-based_2018}.

A more recent definition of ABM suggests that a model should contain a 'sufficient' number of individual heterogeneous entities that posses attributes that are exclusive to themselves and that engage in interaction that alters the states of other entities \citep{polhill_crossing_2019}. Along with the definition, and the listing of benchmark practices and choices in economics, the modeling community has agreed on a support for clear communication of models as well as availability of simulation code\citep{grimm_towards_2014,grimm_odd_2020} \footnote{PolicySpace2 full code is available at GitHub (anonymized)}.

ABM has been used on some real estate market analysis \citep{ge_endogenous_2017,geanakoplos_getting_2012,axtell_agent-based_2014,baptista_macroprudential_2016,huang_review_2014,carstensen_agent-based_2015,goldstein_rethinking_2017,yun_housing_2020}, mostly evaluating macroprudential initiatives towards curbing volatility. One of the first ones is an abstract model for the United Kingdom market proposed by \citet{gilbert_agent-based_2009}. The authors aim at replicating real estate stylized facts, including the role of the broker. Prices are fixed in the short-term and demand is driven by newcomers. The model suggests that lower loan-to-value (LTV) limits curb prices, whereas exogenous demand drives prices up. 

A series of papers focused on the subprime bubble bursting and boom analysis, started with the proposal of \citet{geanakoplos_getting_2012}. \citet{baptista_macroprudential_2016} and then \citet{goldstein_rethinking_2017} develop the model for the Washington, DC case, whereas \citet{axtell_agent-based_2014} apply it to the United Kingdom. The emphasis of Baptista et al. is on learning more about the behavior of investors who buy-to-let, besides discussing the imposition of limits to leveraging. Goldstein studies the influence of the percentage of income that is directed towards the real estate market. All the papers suggest that there is a strong relationship between LTV and the occurrence of more volatility. \citet{carstensen_agent-based_2015} takes the modeling to Denmark in order to investigate effects of shocks on interests and salaries. His work suggests that increases of debt-to-income (DTI) ratios may lead to the collapse of the market. 

\citet{ge_endogenous_2017} adds to the literature discussing volatility in the real estate market, however with a more detailed focus on the bank as an agent that decides on mortgage levels. The bank performs an endogenous calculus that considers the value of collaterals and the probability of default to set mortgage rates. Shocks on the model include the number of investors that act speculatively. She shows that those are sufficient conditions to generate endogenous bubbles in the market. 

Other authors focus on spatial changes and evolution. \citet{prunetti_utility-based_2014} design a utility analysis associated with a land-use and land cover model to represent spatial dynamics. \citet{moeckel_constraints_2017} associates a model of land use change to a transport model to tackle households' simultaneous constraints. \citet{huang_review_2014} review decision-making real estate models that are associated with land-use dynamics. 

PS2 includes a bank that collects clients' deposits, pays interests and offers mortgage loans for prospect buyers. Spatial rules mediated by access to public or private transport are present in the labor market -- as criteria for candidates who are choosing firms --, in the goods and services market as criteria for consumers choosing firms -- along with prices --, and as an influence of properties' price-setting mechanism. Administrative space is also relevant as five different taxes are collected and transferred to the municipalities following tax distribution rules. Households and firms are generated following intraurban census block level data. 

Despite the mortgage mechanisms and the spatial emphasis, we believe a greater contribution of PS2 is the endogenous dynamics of the whole process. Moreover, PS2 enables the analysis of sectorial different policies and their encompassing overall effects. Workers participate in the labor market and may get a job depending on their qualification levels, place of residence and access (or not) to private transport. Employed workers receive salaries and their level of consumption also depends on the household income and its composition (age and employment). The level of demand of households and their location vis-à-vis that of the firms determine how much firms sell, which may vary as the pool of firms that households evaluate changes every month. Firms' sales in turn determine whether the firm is hiring or firing employees. Savings of households dictate how their participation on the real estate market happens. At the end of the month, taxes collected at the various markets are calculated and transferred to municipalities that invest into improving the quality of life of its citizens, via the neighborhood and, thus, properties values. When the policy experiment is in effect, part of the budget is directed towards the chosen policy. 

\subsection{Housing and policies in Brazil}\label{housingBR}
Housing has been a social problem in Brazil since its early industrialization at the beginning of the $20^{th}$ century. Lack of adequate housing and urban sprawl intensified in the 1960s with exodus from rural areas, concentration in cities, and removal of poor housing from central locations. Policies to provide financial support made by the National Housing Bank\footnote{Banco Nacional de Habitação -- BNH, in Portuguese} -- created in 1964 -- mostly helped middle-class households and had its projects overcome by high inflation in the 1980's debt crisis. 

Housing policies were only reintroduced in the 2000s with the creation of the Ministry of Cities. Huge investments in the Program "My House, My Life" (MCMV)\footnote{Minha Casa Minha Vida -- MCMV, in Portuguese.} led to the construction of nearly 5 million housing units within 10 years (2009-2019). However, according to the same government report that totals the production \citep{economiamcmv}, the housing deficit -- an indicator that computes different aspects of housing inadequacy -- remained stable at 6 million units over the period. 

Furthermore, land prices constrained the construction of house units in serviced areas, enabling units in areas which lack infrastructure and access to basic services, as noted early on by \citet{krause2013minha}. On the supply side, the more services and quality of life that are present in a neighborhood, more expensive is the land. On the demand side, 56.25\% of all Brazilian households in 2014 were classified at the lowest bandwidth of the policy \citep{fioravante_credito_2018}. That means that more than half of all households were eligible for the highest subsidy planned in the program, given their insufficiency of income to afford basic housing. 

Since 2019, the policy program has struggled due to financial restraints on federal government expenditure. Hence, access to housing remains as a relevant social issue in need of adequate policy handling. 

\subsubsection{Case study: Brasília Metropolitan Region}

Brasília is the planned capital of Brazil, which was built from scratch and inaugurated in 1960. It is enclosed in the rectangular Federal District, inside the state of Goiás. The conurbed Metropolitan Region of Brasilia includes the 9 neighboring municipalities and it had a population of 3,360,552 inhabitants in 2010 \citep{ibge._ministerio_do_planejamento_arranjos_2015}. 

Housing occupation is sprawled inside the Federal District and the municipalities, generating long distances and high commuting times. \citet{pereira_tempo_2013} estimate an average of 34.8 minutes in 2010. On top of this average, consider that lower income households live further away and do not have access to faster, private transportation means.\footnote{Motorization rate is at 37.3\%} 

In fact, peripheral regions in Brasília, and across metropolitan regions in Brazil, mean poorer access to urban services such as utilities (clean water and sanitation), public security, universal public healthcare and schooling \citep{furtado_territorio_2013}. Metropolitan regions also display high levels of inequality \citep{salata_boletim_2021} with the prominence of the central cities relative to neighboring municipalities \citep{furtado_territorio_2013}. 

\section{Methods}
\label{methods}

PolicySpace2 (PS2) is an economic model that emphasizes regional, municipal, and intraurban spatial elements of the complex real estate market. A market whose dynamic influences are (relatively) not fully understood, although it produces permanent effects on households and the society as a whole. We adapt and extend the original, open-source model \citep{furtado_policyspace:_2018} that in turns follows the tradition of Gaffeo \citep{gaffeo_adaptive_2008} and Lengnick \citep{lengnick_agent-based_2013}. Lengnick is one of the macroeconomic agent-based models family described by Dawid and Gatti \citep{dawid_agent-based_2018}. 

The purpose \citep{edmonds_simulating_2017} of PS2 is to illustrate a potential explanation as to how alternative public investments in housing and monetary aid among citizens impact the economy and inequality in the long-run. Additionally, PS2 is a descriptive model that enables analogies \citep{grimm_odd_2020} among distant facets of analysis, correlating labor productivity to real estate markets or household savings, for instance. Finally, PS2 makes it easy to endogenously reason about the real estate market and policy interventions as an integrated component of the economic system.

The purpose of explanation is verified as a comparison of simulated versus empirical data and the analysis of the policy experiment. The purpose of description and analogy is discussed in the presentation of the sensitivity analysis and the comments on behavior replication. TRACE methodology \citep{grimm_towards_2014} recommends that besides the purpose and answers, the modeler should also provide the target public and the extent to which the model may be expanded. PS2 serves mainly policymakers and academics interested in real estate and economic dynamics. PS2, however, can also be thought of as a platform that contains a wide number of elements. As such, one could detail specific modules of PS2 and use it for further analysis.

We added extensive changes and adaptations to the original model. We included the credit and rental markets. Construction firms now produce endogenous dwellings following profitability, land availability and supply size. Neighborhood effects, supply size and the time the property has been on the market also influence prices. Mortgage loans as well as bargaining make real estate market negotiations more dynamic. Households make decisions on consumption and savings based on their permanent income \citep{dawid_agent-based_2018}. Empirical data follow intraurban information for the year 2010. 

PS2 aims at incorporating most of the influences listed at the real estate markets literature section \ref{real} into a single modeling platform, including a number of endogenous processes in a data generator scheme that includes dynamics and feedback effects. We briefly list these elements before the model full description.
\begin{enumerate}
    \item Uncertainty towards property valuation is assessed locally using limited knowledge by the buyer. Initial listing price reflects size and quality of the property and its endogenous dynamic location influence. Actual transaction price also evaluates the size of current housing offer and buyers' endogenous savings. 
    \item Evaluating the real estate market is an endogenous decision for households that want to change its composition (marriage, migration), but it also happens exogenously \citep{causa_housing_2019}. 
    \item The dynamics of the neighborhood depends on the activities of firms in the vicinity. These dynamics are endogenous and depend on the consumption of households. 
    \item Construction is also endogenous. Firms calculate most profitable regions -- given current prices -- and check their capacity of construction, available land plots and the size of supply to decide on new projects. 
    \item Households' dynamics -- including demographics (aging, mortality and fertility) -- new marriage (endogenous) and migration (exogenous) are present in the market.
    \item Endogenous labor market, along with distance and public and private mobility costs also influence the real estate market. 
\end{enumerate}

\begin{figure}[!t]
    \centering
\includegraphics[width=\textwidth]{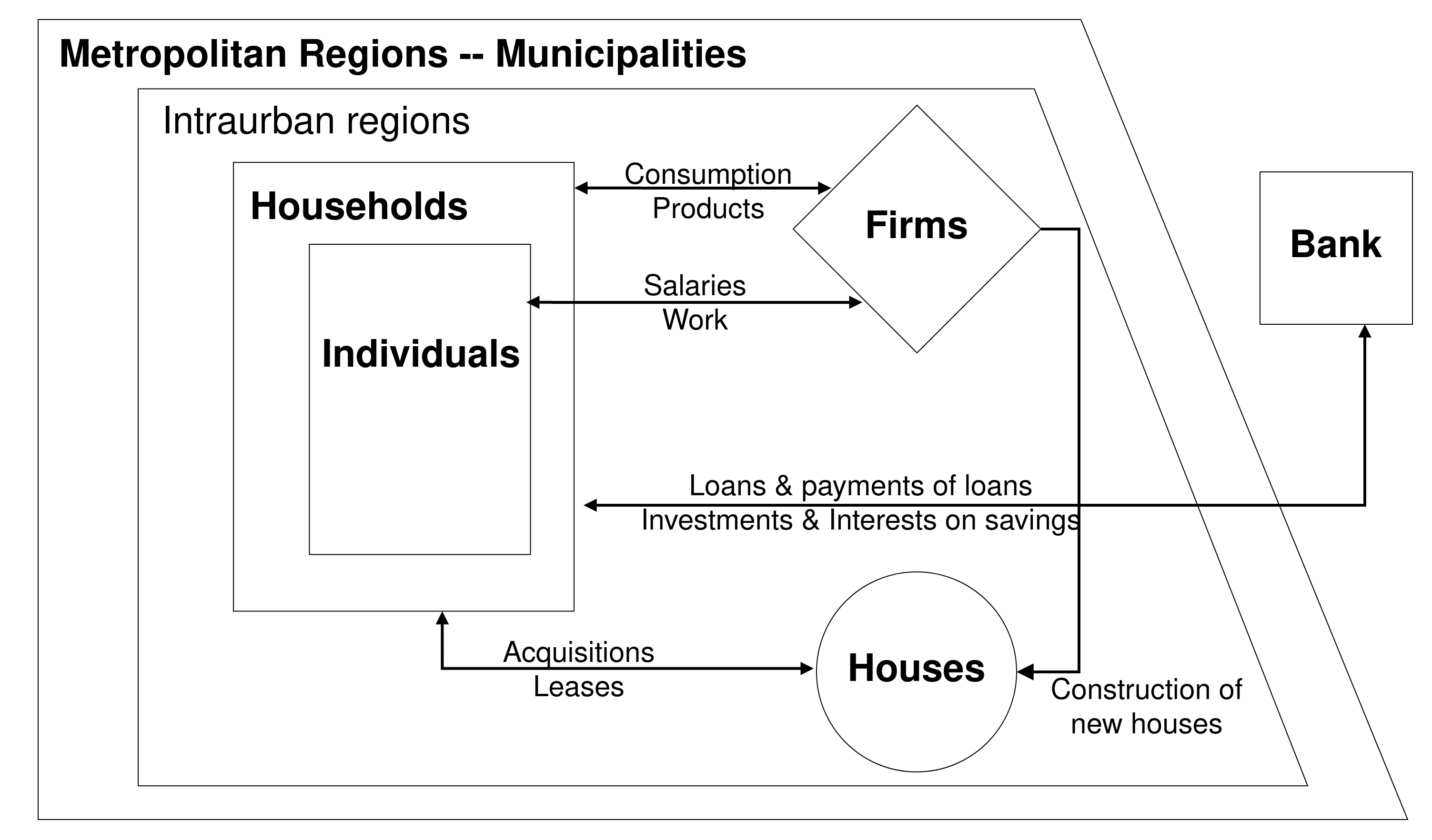}
    \caption{Spatial and macro interactions diagram. PS2 runs individually for each considered metropolitan region (MR). Each MR contains the main official conurbed municipalities. Municipalities are spatially divided by intraurban regions that contain official sampled population data for 2010. Houses and firms are located inside intraurban regions. Individuals are grouped together inside households. They participate independently in the labor market, but make consumption decisions collectively as a household. The bank has no spatial definition. The main interactions occur among firms and households (goods and services markets), individual agents and firms (labor market), among households and households and construction firms (real estate market), and the bank and households (credit market).}
    \label{fig:space}
\end{figure}

We detail the model providing the context of agents and scale and then we use the sequence of events to describe the decision-making processes, the related equations, and the supporting literature. Figure \ref{fig:space} depicts the agents of the model within their spatial hierarchy and the main markets. Each simulation is performed at the level of metropolitan regions. The policy test runs for the metropolitan region of Brasília and its 9 surrounding municipalities. Brasília and each municipality is divided by its component intraurban areas -- officially census blocks called "weighted areas".\footnote{Weighted areas are referred to as "Áreas de Ponderação" in Portuguese by the the Brazilian Statistics Bureau (IBGE).} Houses and firms are located proportionally inside intraurban regions and interact in the markets of labor, goods and services, real estate and credit.

\textbf{Input data.} The data that feeds PS2 comes from official sources and refers basically to the (a) details, location of individuals and households, (b) location of firms, (c) geographic files of municipalities and intraurban regions, (d) state-level data on fertility, mortality, by gender and age, (e) population estimates, human development index, and taxes transferred to municipalities, and (f) general indicators on marriage and car ownership. A full table with sources and descriptive statistics is presented as an Appendix (see Table \ref{tab:input_data}). It is relevant to point out that no data associated to the real estate market enters the model. That means that size, quality, nor prices of dwellings enter the model. All input data is available in the model's repository.

\textbf{Agents.} PS2 contains individuals who work, commute, age, die and are born, get married, and divorce. Individuals are organized in households (families) and reside in dwellings that have fixed locations. Households may move among residences and are considered as a collective of individuals making decisions on consumption. Firms -- also fixed in space -- hire individuals, produce and participate in the goods and services market. Construction firms also hire individuals and supply new dwellings in the real estate market. There is a bank that collects deposits, pays interests on them and offers mortgage loans. Municipalities have actual geographical coordinates, collect taxes on firms and workers, consumption, properties and transactions within their own territory. \\
The model is stock-flow consistent. Workers only receive money from firms. Firms' revenue comes from sales. Households receive payments from rental and real estate markets paid from other households. The bank makes money from loans and pays interests. Municipalities invest only money collected by taxes.

\textbf{Scale.} PS2 runs monthly from January 2010 to January 2020 in the standard simulation. It may be configured to go up to 2030 and start either in 2000 or 2010. Each simulation is run for a single metropolitan region. We consider the urban core of metropolitan regions, defined and named by the Statistics Bureau as Areas of Concentration of Population (ACPs). The model contains input data for all 46 ACPs of Brazil the model.\footnote{Please, check the repository for a full list.} Our study case of Brasília, which is the default run, involves the Federal District (Brasília) and its nine neighboring municipalities.

\subsection{Sequence of events and details}

\begin{figure}[!t]
    \centering
\includegraphics[width=\textwidth]{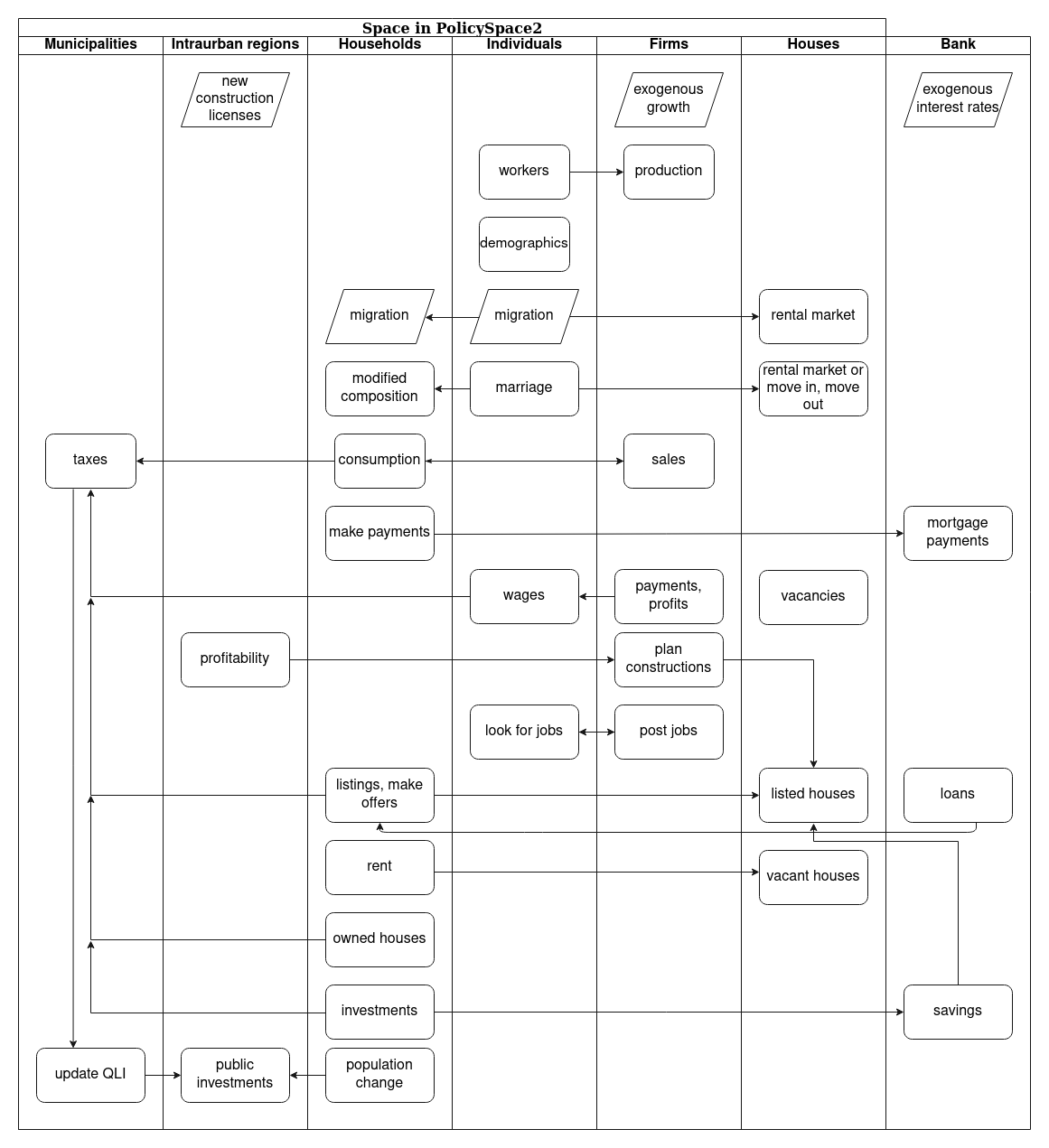}
    \caption{Diagram of macro-processes. PolicySpace2 is a computational discrete agent-based model. The diagram shows major events and interactions as they happen, monthly, from top to bottom. First some exogenous data is computed, then hired workers engage in production. Individual agents age, die and give birth. Migration is processed and new households enter the metropolitan region via the rental market. Marriage and demographics may change household composition, which also might influence the rental market. Consumption and sales are processed, and first taxes are computed. Firms calculate revenues, pay employees and taxes. Construction firms observe vacancies, calculate profitability in different regions and decide on additional construction. The real estate market is operated, mediated via possible loan credits. Households invest. Finally, municipalities balance taxes received and make public investments.}
    \label{fig:sequence}
\end{figure}

Figure \ref{fig:sequence} shows a broad sequence of monthly events that are detailed below. Some decision-making involves more than one group of agents. At the beginning of every month exogenous data such as number of new firms and the mortgage interest rate are computed. Then workers engage in production and individual demographics processes occur. Migration is processed following volume of observed exogenous data. However, when entering the model households have to go through the rental market and successfully get a house. Marriage is relevant as it may change internal composition of households and generate endogenous demand for new houses. Next, households consume, firms processes sales and compute consumption taxes. Firms process payments and pay wages. Construction firms calculate most profitable feasible intraurban regions and decide on construction. Then, the labor and real estate market take place, the latter being mediated by possible mortgage loans. Then, we have the rental market, collection of taxes on real estate transactions, house property, and savings interests paid by the bank to households. Finally, collected taxes are turned into public investments weighted by population change.

\begin{enumerate}
    \item \textit{Generator of agents.} The model either loads previously saved agents or generates them from official census block data. We also use official intraurban geographically delineated regions. Municipalities are a set of regions. Dwellings are generated so that there is an exogenously assigned number of vacant residences in the model \citep{nadalin_empty_2016}. Agents, households and dwellings follow each census block's population percentage ($\text{pop}$) for a given starting year. All houses of the model are allocated to households, thus all houses have owners that are households. The distribution is made so that some households have two houses or more and some have none. Agents are allocated into households and households into dwellings, either as an owner or a rental, randomly distributed.\footnote{Parameters and their standard value are described on Appendix \ref{appendix}.}
    \item Each region lists the number of available land plots or licenses. Construction firms need to purchase a license in order to build a new property.
    \item Following exogenous empirical data, new firms enter the simulation.
    \item Firms produce a homogeneous product using fixed technology \citep{lengnick_agent-based_2013}. They ($i$) update their inventory with goods and services ($Q_i$) each month ($t$) based on workers ($l$) qualification ($q$) \citep{gaffeo_adaptive_2008} and two exogenous parameters ($\alpha,\beta$) (see equation \ref{eq:production}). Construction firms produce dwellings with varying characteristics and location, observing the maximum profitability. However, their production mechanism is the same. 
    \begin{equation}\label{eq:production}
        Q_{i,t} = \sum_{l}^{L} \frac{q_l^{\alpha}}{\beta}
    \end{equation}
    \item \textit{Demographics.} Mortality, fertility, new marriages and aging take place according to exogenous probabilistic official data by state. Each agent receives a month of birthday in which all demographic processes occur. When female agents within fertile age  (15-49) give birth, the child is incorporated in the mother's household. Individuals are relevant in the composition and modification of households and act alone at the labor market. All consumption decisions, however, are made collectively at the household level (see Figure \ref{fig:space}). 
    \item \textit{Migration and marriage.} Migration occurs when necessary to maintain exogenously observed population growth. Households coming into the municipality only stay if they are able to find residence through the real estate market. \\
    Marriage occurs probabilistically. Agents leave their old household (and house property) behind, if not the only adult. Otherwise, they bring the children (or property) with them, if they have any. Newly formed households persist only if either adult brings a house (or a rental) or if they succeed in the market. \\ Inheritance. When the last member of a household dies, a search for relatives occur. Members from the household that the deceased originally belonged to receive any wealth or property. When there are no known relatives, any owned property is randomly allocated to another household.  
    \item \textit{Consumption at the goods and services market.} Households choose from an exogenously determined sample size ($\varsigma$) either the firm that is the closest or the one with the cheapest product ($P(.5)$). The consumption amount is determined by the household's simplified permanent income ($PI_{h,t}$) \citep{dawid_agent-based_2018}, with an extra assumption that expected future income is an average of all previous permanent income (see equation \ref{eq:permanent_income}).\footnote{Permanent income is "a linear function of current and expected future incomes and of financial wealth." \citep[p.78]{dawid_agent-based_2018}} When gathering consumption money, households search first for cash available with each member. If collected cash is not enough to make the permanent income, households try to withdraw from their reserve money ($R_h$) or savings ($S_h$) from the bank. \\
    The reserve money is simply some cash kept to accommodate fluctuations of wages and balances, which remains in possession of the household and is not invested in the bank. It provides immediate cash for payment of rentals, loans and consumption. The distinction is only a household internal separation. Reserve money is not invested in the bank, thus receives no interest. It is given as six times the Permanent Income: $R_h = 6*PI_{h,t}$.
    \begin{equation}\label{eq:permanent_income}
        PI_{h,t} = i_t * \overline{Y_{h,t0-t}} + i_t * \frac{\overline{Y_{h,t0-t}}}{r_t} + w_t * r_t 
    \end{equation}
    where $i_t$ is $r_t/(1+r_t)$ and $r_t$ is the baseline interest of the economy, $\overline{Y_{h,t0-t}}$ is the household's members average monthly income for all available periods and $w_t$ is the sum of property values, reserve money, savings and loans. Any income the household may have (or gain via rentals or sales) in excess of permanent income is deposited as savings in the bank, but for the reserve money. Simply put, households consume a bit more than total wages if they have savings and a bit less when their total wealth is negative. \\
    Note that consumption happens before firms have paid wages in the current month. Thus, households use resources from last month and their reserve money. In the sequence that follows, wages are paid and monthly housing obligations, that is, rents or mortgage payments, are fulfilled. When the household has no funds from wages, reserve money or savings, it goes with null consumption for that given month. Null consumption is very low for the default configuration of the model. Given the context of Brazilian current status in 2010s, these numbers are compatible with observed data. At the end of the month, households check whether saving investments are possible. Household savings enable the offers on real estate market.
    \item The bank calculates and collects payment for loans. If the household fails to pay in full, the debt accumulates for the next month. Note that the bank pays exogenous baseline economy interest on deposits ($r_t$), but applies (also exogenously) market mortgage interest rates on loans.\footnote{See \citet{ge_endogenous_2017} for an example of an endogenous mortgage rate mechanism.} We use real observed mortgage rates as default. We also tested nominal and fixed rates at the sensitivity analysis. 
    \item Firms check their revenue, pay taxes and calculate profit. Calculate wages, pay their employees and decide whether to update prices. \\ 
    \textit{Firms' decision on prices.} \citet{blinder_sticky_1994} identifies via survey a number of different practices firms use when setting prices under uncertainty. His findings support the idea that firms do not evaluate the market every month. PS2 follows \citet{seppecher_what_2017} in observing the size of the firms' inventory in order to update prices and it does so not every month, but according to an exogenous parameter ($\zeta$) \citep{hamill_agent-based_2016}. If the amount sold in the previous month was above produced quantity, then firms update prices by a markup percentage ($\pi$).\\
    \textit{Firms' decision on wages.} Firms decide on wages ($\omega_{l,t}$) based on total revenue ($TR_i$) discounted of taxes on labor ($tax_l$) and global unemployment ($U_t$) (see equation \ref{eq:wages}). In practice, workers' payments vary according to the firms' sales and represent their own contributions towards production. The rationale may be considered as a variable bonus attached to firms' performance.  
    \begin{equation}\label{eq:wages}
        \omega_{l,t} = TR_{i,t} * (1 - U_t) * \frac{q^{\alpha}_{l}}{\sum_{l}^{L} q^\alpha} * (1 - tax_l)
    \end{equation}
    \item \textit{Planning new dwellings.} Construction firms operate their property planning process considering availability of land plots, profitability, and current supply size. They also check for finished previous construction plans and when so list the new properties in the market. First, the firm checks whether their contracted amount of work ($\sum_h cost$) is smaller than a fixed number of months ($n$) times their current monthly production ($Q_{i,t}$). If so, they may start a new construction project. Next, the construction firm checks whether it has enough funds to buy the license plots and separates the regions where it can afford and that actually have available licenses. The firm then tests whether to continue with the construction plans checking probabilistically against global vacancy percentage. That means that the higher the offer on the market the less likely the construction firms start new projects. \\
    In order to evaluate profitability in the regions where there are licenses available, a planned building size and quality are chosen randomly. Next, the firm gets a sampled price of houses that have similar size (within 10 absolute distance) and quality (within 1 absolute distance) for each available region. Cost is calculated as dwelling size ($H_s$) times quality ($H_q$), times a random productivity factor that is a function of markup ($f(\pi)$). Region profitability ($N_{\pi,m}$) then is the mean prices of similar dwellings deduced by calculated building cost and license price ($N_m$), times lot cost ($1 + \upsilon$) (see equation \ref{eq:construction_profit}). Given this planning information, the firm evaluates which plan has the most profitability and makes the decision to build at the place and characteristics where it has the highest profit, relative to current mean prices of similar houses. When there are no profitable regions, the firm does not start a new construction. 
    \begin{equation}\label{eq:construction_profit}
        N_{\pi,m,t} = \overline{P_{\text{ask},m,t}} - (H_s * H_q * f(\pi) * N_{m,t} * (1 + \upsilon))
    \end{equation}
    \item \textit{Labor market.} \citet{neugart_agent-based_2012} review an initial ABM labor market. A full scale labor market is describe at \citet{axtell_endogenous_2013}. They are able to reproduce a number of labor market stylized facts. In PS2 we have opted for a simple labor market \citep{hamill_agent-based_2016}. All individuals ($l$) who are between 16 and 70 years old and out of the job market apply every month. Consumption decisions are all made at the level of the household and the model does not include a pension system, nor unemployment benefits. Brazilian social security rules is complex and varies widely according to employees' trajectories, sectors, and whether the worker belonged to government or private institutions. Moreover, pension and unemployment benefits cover only formal labor (informal labor accounts for nearly 50\% of the population), and are small in value. \\
    Firms ($i$) do not enter the market offering positions every month. Rather they evaluate probabilistically whether to do so following an exogenous parameter ($\iota$). When in the market, they may either fire a randomly chosen employee, when their profit is negative, or open a position otherwise. Candidates and firms are shuffled. Depending on an exogenous parameter ($\eta$), some of the posts available will use a spatial proximity criterion whereas others will make a decision based on qualification of candidates. \\
    The spatial proximity makes sense for the case of Brazilian cities where legislation imposes that transport costs should not exceed 6\% of wages, with employers bearing values above this limit. Consider further that populous Brazilian metropolis are spread across large areas, with low-quality commuting infrastructure, thus making traveling large distances to work costly for both workers and employers. \\
    All positions involve a sample of candidates ($\sigma$). The candidates at the pool for each position evaluate the post themselves considering the base wage ($\Omega$) paid and the distance from each firm to their residence ($d_{h-i}$), mediated by the cost of transport ($c_{tr}$). Having access to private transport depends probabilistically on the deciles of income of the candidate's last wage. That is, higher-income households in Brazil typically owns a vehicle and evaluates time commuting more heavily. As such, those candidates prefer positions that are closer to their residence. Candidates on lower quantiles of income  who have no alternative but public transport also penalizes distance, however, at a lower comparative scale. Although distance and transport are considered as criteria at the labor market, household transportation costs are not included in the model.\\
    Firms' base wage ($\Omega$) is the total amount distributed by the firm on a given month and functions as a \textit{proxy} that informs the candidate about the size of the firm, which is usually associated with better career paths, salaries and stability. It seems there is no consensus about the firms' wage decision-making process \citep{neugart_agent-based_2012}. Additionally, each firm classifies candidates by their qualification ($q_l$). All offers are sorted based on the score that considers both candidates and firm parameters (see equation \ref{eq:score}). When firms choose based on proximity exclusively, according to ($\eta$), then candidates' qualification ($q_l$) is not considered in the computation of score ($s_{l,i,t}$). Firms paying higher wages choose first in descending order. For every pair of firm-candidate -- conditional on the candidate having not being chosen earlier -- positions are filled successively and firm and candidate leave the market for that month. The market closes when there are no more positions or no more candidates. \\
    \begin{equation}\label{eq:score}
        s_{l,i,t} = q_l + \Omega_{i,t} - d_{l,h-i,t} * c^{tr}_{l,t} 
    \end{equation}
    \item \textit{Real estate market.} Entering the real estate market is endogenous due to changes in household composition (marriage, divorce) and migration. Households also monthly enter the market exogenously, according to a sample of households ($\phi$). However, entering the market does not necessarily means making a new purchase or renting a new house. The offer may not be accepted, the bank may decline the loan, or the new rental may not be better than current dwelling. Houses are listed in the market when vacant. \\
    Prices ($P_\text{ask}$) for all properties are updated and those unoccupied are listed and divided between rental and sales market by an exogenous parameter. That means that all unoccupied houses are available in the market. Affordability of both rental and purchasing is the main factor when choosing a house in Brazil. As noted above, more than half of the households would be eligible for federal housing program at the highest level of subsidies. \\
    \textit{Rental.} In addition to the rental market transactions that might have occurred during migration and marriage procedures, rental market happens first when the main real estate market is run. Households participating in the rental market are sorted by permanent income ($PI_{h,t}$). Rentals are initially calculated as a fixed exogenous percentage of the house calculated value. Given a random sample of fixed size ($\sigma$), households ($h$) choose a rental that is within their budget at price value. When no house that the household can afford is available, the household propose a discounted value on the cheapest rental in the sample. When negotiations are not accepted, households remain at their current house and leave the market. Even when evaluating the market positively, households do not conclude the transaction when their current house is better than the prospective one.\\
    \textit{Sales.} On the sales market, households are sorted by purchasing power ($P_{\text{offer}}$), including an estimate of possible mortgage credit ($L_h$). Then, each household tries to buy the best house on their sample \citep{goldstein_rethinking_2017}, considering size, quality, neighborhood service levels, neighborhood households income, and time on the market. All these elements together compose the price. \\
    \textit{Asking prices.} The asking price is calculated considering the dwelling size, quality ($H_{s,q}$), and the neighborhood quality of life index ($N$) which differs by region ($m$) and changes monthly depending on taxes invested and population proportion (see below item 15)\citep{rosenthal_change_2015}. This follows the hedonic price literature that decomposes house value as the cumulative sum of its characteristics \citep{rosen_hedonic_1974,malpezzi_hedonic_2002}. Additionally, an extra comparative effect of neighborhood -- also testes in the sensitivity analysis -- is added in the default configuration. A parameter ($\tau$) brings the influence of a normalized index of neighborhood households average income ($N_q\in[0,1]$) into prices \citep{ge_endogenous_2017}. Finally, a discount for time on the market is incorporated into prices with a bounded value ($\gamma$) and a decay factor ($\kappa$), depending on the number of months the property has been listed ($T$) (see equation \eqref{eq:p_ask}).\\
    \textit{Bank mortgage criteria.} The bank follows three simple criteria to provide mortgage loans: (a) the bank needs to have a positive balance; (b) the household cannot have a current mortgage, and (c) total loans already offered cannot be higher than the percentage of total deposits ($\nu$) as established by monetary authorities. When those criteria are valid, the amount ($L_h$) the bank provides depends on the maximum capacity of monthly payment the household can make, restricted by a limit ($\chi$) of permanent income ($PI_h * \chi$) times the maximum number of months ($m$) (360) or the number of months before the oldest member of the family reaches 75 years old, so that $L_h = PI_h * \chi * m$  \\
    \textit{Negotiation.} When the property's price is below the household's savings, the transaction is made with final price ($P$) set as a simple average of asked price and household's savings \citep{dweck_discussing_2020}. When property's price requires mortgage loan ($L_h$), the buyer requests the loan on the difference between savings and asked price. If successful in getting the mortgage loan, the offer price ($P_{\text{offer}}$) is savings plus estimate mortgage (see equations \eqref{eq:offer} and \eqref{eq:p}). Otherwise, when the loan is declined by the bank, the household leaves the market. When the savings are below the asked price, but above an exogenous parameter limit ($\rho_{-}$), the buyer makes an offer with their total savings. The chance that the seller will accept depends probabilistically on the size of the supply market (see equation \eqref{eq:reduced}). When both savings and savings with mortgage are below property asked price, and the discount was not possible or not accepted, households try the next house on the list. \\
    \textit{Constraints.} There are some constraints imposed on prices. The ratio savings, price is upper-bounded by an exogenous parameter ($\rho_{+}$) and there is a loan-to-value parameter ($\text{LTV}$) on mortgage requests (see equations \eqref{eq:c1} and \eqref{eq:c2}).\\
    \textit{Rental and mortgage payments} In the case that the household does not have enough resources to pay for rent, tenants default on rent and landlords bear the loss. These defaults on rent are very low on a typical simulation run and are compatible with empirical data. When households default on mortgages, the bank incorporates the temporary loss and tries to recover it in posterior months.

\begin{gather}
    P_{\text{ask}} = H_{s,q} * N_{m,t} * (1 + \tau * N_q) * ((1 - \gamma) * e^{\kappa * T} + \gamma) \label{eq:p_ask}\\
    P_{\text{offer}} = S_h \lor S_h + L_h \text{ if } P_{\text{ask}} < P_{\text{offer}} \label{eq:offer}\\
    P = ( P_{\text{ask}} + P_{\text{offer}}) /2 \label{eq:p} \\
    L_h / P <= \text{LTV}  \label{eq:c1}\\
    \text{if } P_{\text{ask}} / P_{\text{offer}} > \rho_{+} \longrightarrow P = P_{\text{offer}} * \rho_{+} / 2 \label{eq:c2}\\
    \text{if } P_{\text{ask}} > S_h > \rho_{-} \longrightarrow P = S_h | P(\sum \text{Listed} / \sum h) \label{eq:reduced}
\end{gather}

    \item \textit{Decision on moving.} \citet{jordan_agent-based_2012} list seven reasons why households change residences. Those include changes in the socioeconomic status (such as losing jobs), and searching for better quality of life in a better neighborhood. As a decision made by the household unit, the financial power of its members influences the capacity to acquire larger and better houses. Households will move to the best dwelling \citep{goldstein_rethinking_2017}) when at least one member is employed. In the rare case that all adult members are unemployed, households will move to the worst house they own and sell the best one in order to get capitalized and adjust their familial budget. Moving occurs endogenously within PS2 when a household status change, due to marriage, for example, or a migration event. Exogenously, the household will at times participate in the real estate market, but the purchasing and moving only happens when the full process occurs successfully.
    \item Households invest. The bank keeps the date to calculate interest at the exogenous rate ($i_t$).
    \item Municipalities invest in Quality of Life Index improvement ($N_{m,t}$). All taxes collected: consumption, labor, firm profits, house transactions and house taxes, are transferred to the municipalities budget according to tax rules distributions in Brazil, following the original model \citep{furtado_policyspace:_2018}. Investments are linear and transformed via an exogenous municipal efficiency index ($\psi$) and population ($\text{pop}_m$) difference, so that: $N_{m,t} += \sum_{m} tax_t * \psi * \text{pop}_{m,t-1}/\text{pop}_{m,t}$. The index is thought of as a \textit{proxy} for the Municipal Human Development Index (M-HDI) and it represents the municipal capacity to collect taxes -- given by the dynamics of markets within its boundaries -- weighted by the proportion of its population. Hence, areas that have markets that are proportionally more dynamic invest more in infrastructure and amenities. All municipalities start with their observed M-HDI in 2010. $\psi$ is calibrated so that values maintain a resemblance with M-HDI. Municipal investments (or lack thereof) impact house prices as they reflect these changes in urban quality. Implicitly, households living in places with higher QLI enjoy a better level of services and amenities.
    \item Statistics and output are calculated and saved. Gini coefficient is based on households' permanent income (see equation \ref{eq:permanent_income}). That means that average household income as well as housing property or mortgages enter the calculation. GDP is calculated as the sum of firms' revenues by municipality and aggregated for the metropolitan region.
\end{enumerate}

\subsection{Policy experimental design}
\label{policy}

The policy experiment is applied using endogenously collected taxes by municipality ($m$). Instead of investing full budget to the improvement in quality of life ($\sum_{m} tax$), a percentage ($\delta$) is diverted to either one of three policy experiments (Table \ref{tab:policy_experiments}). This endogenous mechanism guarantees a high degree of comparability among the three tested policies.

When applying a policy, the first action is to register all households whose permanent income ($PI_{h,t}$) has been below an exogenous quantile ($\theta$) of the households of the metropolitan region in the previous year and currently resides within the municipality. Households are then sorted according to permanent income so that the poorest of all municipal households is the first in line (see Table \ref{tab:policy_experiments}). 

\begin{enumerate}
    \item \textit{Property acquisition and distribution.} The municipality lists all properties in the region that construction firms have finished but have not sold yet and sorts them by the cheapest to the most expensive. The list of households registered includes only those who do not own any property. Then, the municipality purchases the property from the selling firm, and transfers the property to the household next on the list. The municipality buys and transfers houses for as long as the monthly available allotted funding, properties, and households last. Benefited households do not pay for the houses received. 
    \item \textit{Rental payment vouchers.} The municipality lists all households that do not own any properties and are in the policy register. Thus, as long as there is enough funding and households the municipality issues 24-month rental vouchers that should cover the current household rental price. Rental vouchers are attached to the current rental contract. If the household decides to leave the residence, it gives the remaining vouchers, if any, back to the municipality. Households can only apply for new vouchers after they have expended all previously received and the criteria to be listed still hold. 
    \item \textit{Monetary aid.} In this policy scenario, the municipality divides all monthly available funding equally into a single payment for all households listed at that moment in the policy register. Households are free to use the received funds. 
    \item \textit{No-policy -- baseline.} In this case, no money is invested in policy and all resources go into investment in municipal quality of life ($N_m$). 
\end{enumerate}

\begin{table}[!t]
\centering
\begin{tabular}{lcll}
\toprule
 Policies & $\delta$ portion of municipal budget $\sum_{m} tax$ &
 Benefit period &
 Amount received \\
  \midrule
Property acquisition  & .2 & Permanent & The house itself   \\
Rental payment vouchers               & .2 &  24 months     & Rental value \\
Monetary aid                          & .2 &  Current month & Variable  \\
No-policy baseline                    & 0   &    &      \\
\bottomrule 
\end{tabular}
\caption{Comparative analysis of policy experiments. All policies tested use of proportion ($\delta$) of an endogenous collection and transfers of municipality taxes ($\sum_{m} tax$). At the no-policy baseline scenario, all of the endogenous budget go to the general public investment of the model. The house received by the Property acquisition policy is permanently transferred to the household. Rental vouchers are given for a maximum of 24-month period. If the household decides to leave the rental, the benefit is lost. The amount received by households in the Monetary aid policy depends on the amount available for investment and the number of households that match the criteria: permanent income ($PI_{h,t}$) below an exogenous quantile ($\theta$) of all households.}
\label{tab:policy_experiments}
\end{table}

\section{Verification, calibration and sensitivity analysis, and validation}
\label{val}

Computational analyses are always subject to possible errors of execution or actions that differ from the original intention of the modeler \citep{galan_errors_2009}. In order to avoid the occurrence of such errors, programmers need to verify that the code executes as intended. Some procedures applied to PS2 aimed at warranting processes ran as expected. 

We observed 66 different outputs of the model to precisely follow results and changes due to altering of parameters, procedures or code. Moreover, we included a series of \texttt{assert} commands throughout the code -- which is available at a public repository -- to make sure of the values of variables at key points. Finally, we ran specific tests to check, for example, whether construction firms actually build new houses, banks effectively loan to families or whether there is any household without a current address. Specifically, we tested the flow of financial resources among agents to ensure that the model was cash-flow consistent.

We aim at validating PolicySpace2 through a series of successive steps. First a series of macroeconomics indicators have to behave reasonably within expected values. Even though these indicators have been calibrated to be in such a way, they happen to be within reasonable boundaries simultaneously. That means the Gini coefficient, inflation, and unemployment among other indicators are all sensible. 

Brasília -- which is our baseline case -- observes Gini coefficient of 0.4705, total inflation for the 10-year period is 43.32\%, and unemployment 12.39\% which are within expected values for the case of Brazil. Apart from that, extensive variations in the parameters (or the metropolitan regions) result in different values but do not lead to the collapse or exponential behavior of the results of the model. 

Most rules, procedures, and parameters come from literature or data. Firms' decisions on prices, wages, and production, for example, are based on previous works. Price setting in the real estate market follows hedonic and urban economics baselines with also some support from urban studies. Labor market does not have a clear predecessor, despite the contributions of \citet{neugart_agent-based_2012}, but it has based itself on commuting costs and activities' time allocation. Parameters follow observed data as much as possible. 

Moreover, a number of new rules implemented in PolicySpace2 are tested in the sense that they can easily be turned on and off. Specifically, we tested three structural rules and investigated eight different aspects of the model in order to comprehend further the underlying mechanisms. We also performed sensitivity analysis on the models parameters.

\begin{figure}[!t]
    \centering
\includegraphics[width=\textwidth]{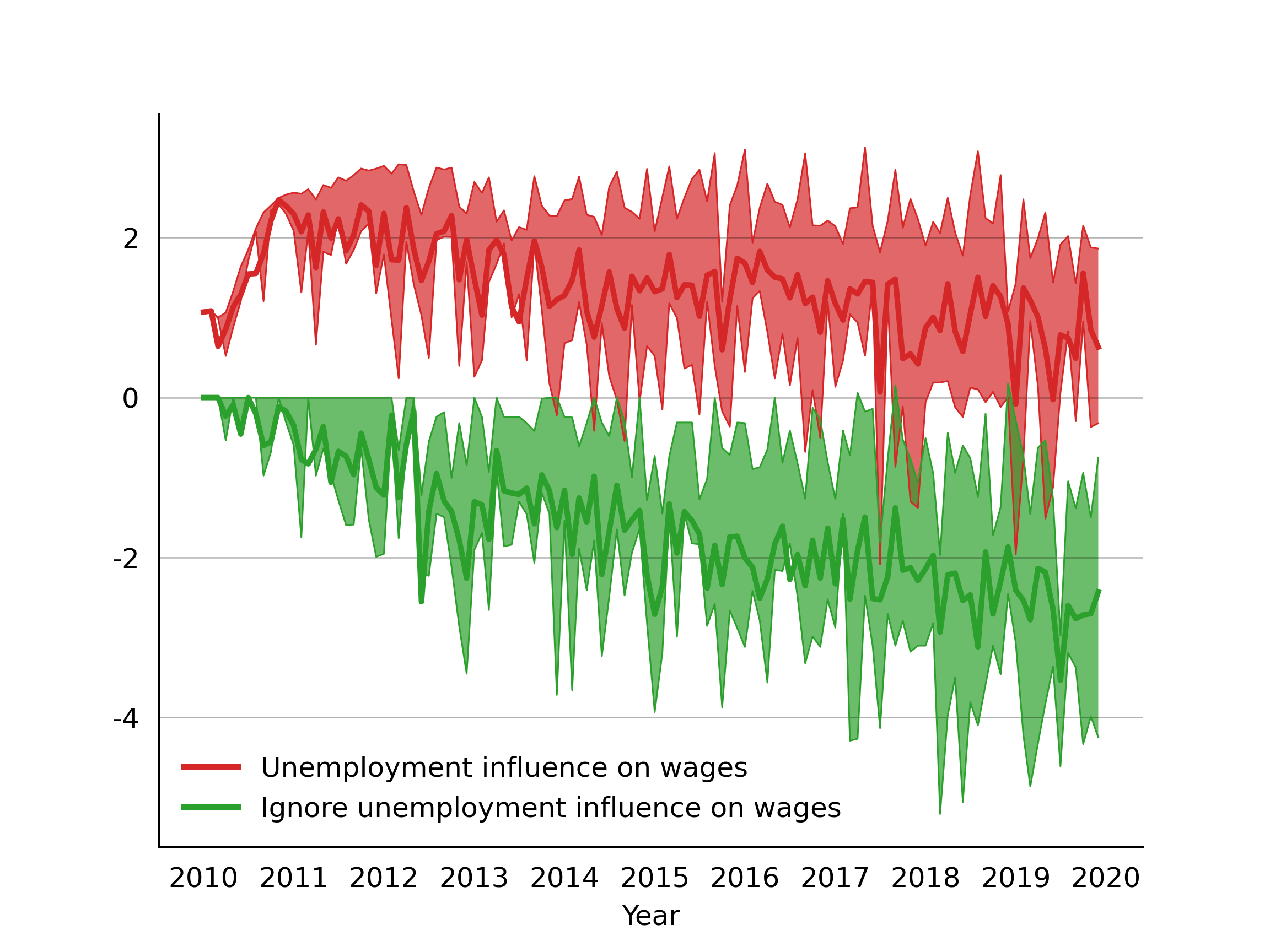}
    \caption{Average profit of firms, superior and inferior quartile, for 20 runs with each setting of global unemployment ($U$) for the case of Brasília 2010--2020. Red refers to the simulations when $U$ influences-- firms' wage-setting. Green represents the simulations when the rule is absent. Discounting wages proportionally to unemployment provides firms with additional resources to handle demand volatility, thus maintaining a positive trajectory and balanced results.}
    \label{fig:profit}
\end{figure}

\begin{enumerate}
    \item Proximity to labor market ($\eta$). We tested the influence of $\eta$ from 0 to 1. When $\eta==0$ distance is not considered in the labor market and all candidates are selected exclusively based on their qualification. When $\eta=1$, candidates are selected based on distance only. The analysis suggests that most values between the extremes are reasonable in the sense that unemployment and the other indicators remain stable. When $\eta$ is set in the extremes of 0 or 1, unemployment is much larger, GPD is lower, and total household commuting is much lower.  
    \item Neighborhood effect on house prices ($\tau$). We tested $\tau$ in the interval 0 to 5. When $\tau=0$ the rule is not in effect and the neighborhood average income -- a \textit{proxy} for neighborhood value and perception -- is not included in price calculation. As a result, prices are lower and GDP is higher. Perception of neighborhood value for real estate prices, however, is relevant for price composition \citep{8631423}, thus our default value $\tau=3$ includes a moderate influence of the neighborhood in prices. 
    \item Global unemployment  ($U_t$) as a factor on firms' wage setting. We tested the presence or absence of unemployment on firms' mechanism of wage-setting. Mostly the presence of the rule influences the finances of the firm providing it with additional money availability that enables the firms to counteract the high volatility of demand observed in the model (see figure \ref{fig:profit}).   
\end{enumerate}

After verifying the code and calibrating the parameters, a thorough sensitivity analysis was applied to the final model. The sensitivity analysis served the purpose of comprehending the mechanisms behind the model and the interactions across different markets and agents. In total, we ran 5,573 simulations with more than 244 unique configurations of parameters. All parameters are tested with a variation of combinations. That helped verify the robustness of the results as well as to gain insights on the operational emergence of results of the model.\footnote{Please note that PS2 comes with a 'sensitivity' run that automatically tests Boolean, quantitative parameters, tax rules distribution, a run with all metropolitan regions, and all policies. The code encloses built-in runs that provide output as comparative plots.} Specifically, we tested all five taxes with different values for the parameters. The results showed different levels of magnitude for the results, however, without changing the typical trend of indicators.

Considering the purpose of the model -- that is to investigate "alternative household poverty alleviating mechanisms", we tested the policy experiment in four other metropolitan regions. All tested rules and parameters maintained the bulk of the results presented with similar observed trends (see \ref{tab:comparison_among_cities}). 

Validation itself is done by comparison of data collected for the Brasília real estate market that never goes into the model. It was gathered independently from listing offers available on internet sites mostly during 2020. The empirical data contains 23,103 observations across 61 neighborhoods with a median house selling for R\$ 750,000 with three bedrooms in 126 squared meters and R\$ 6,011 per squared meter. Model data, however, only uses information from Census and official data, mostly from 2010 (although it is also possible to run with 2000s data). 

\begin{figure}[!t]
\centering
\subfloat[Empirical house prices data.]              {\includegraphics[width=.61\textwidth]{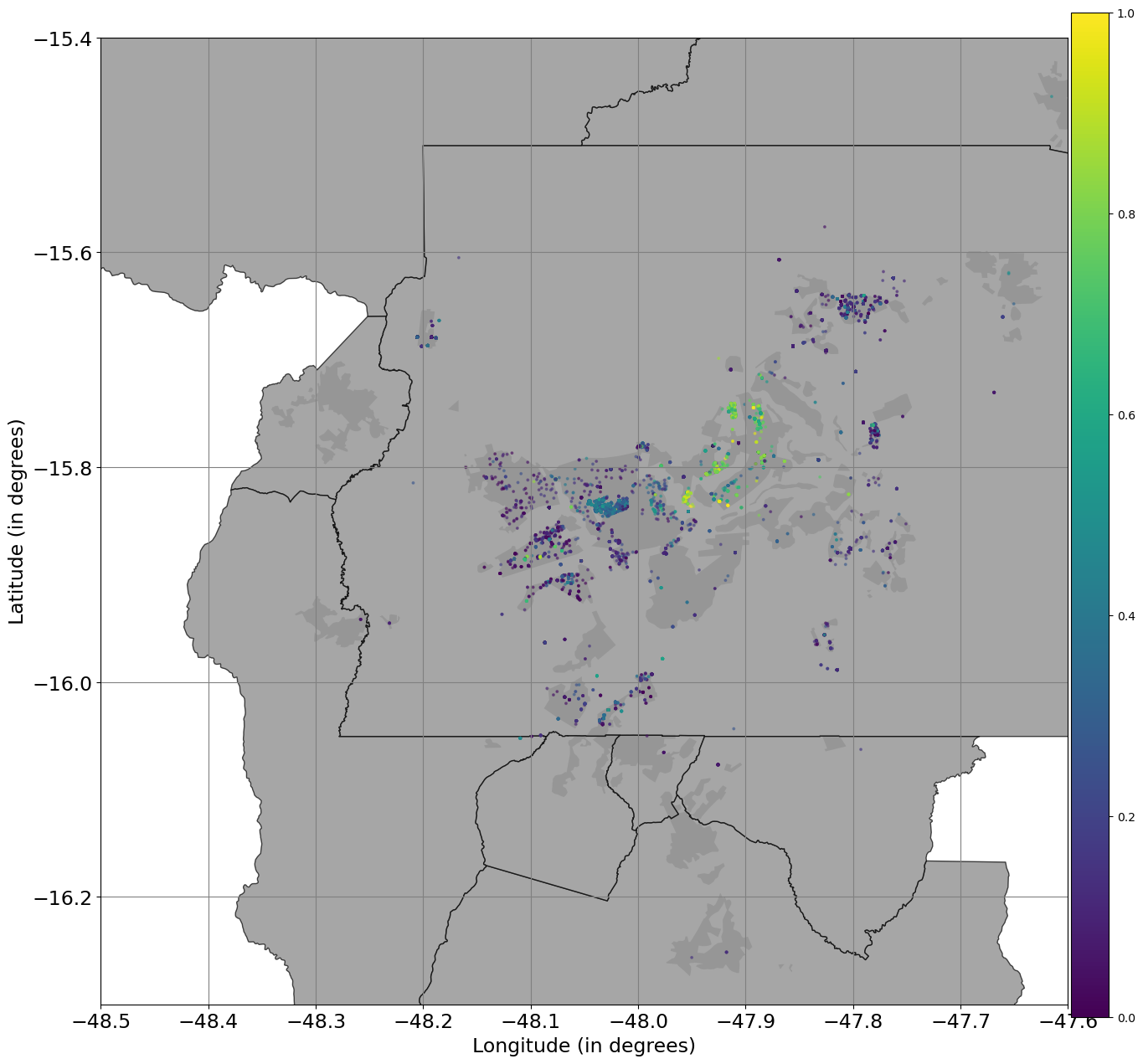}}\qquad
\subfloat[Simulated house prices data -- No-policy.] {\includegraphics[width=.61\textwidth]{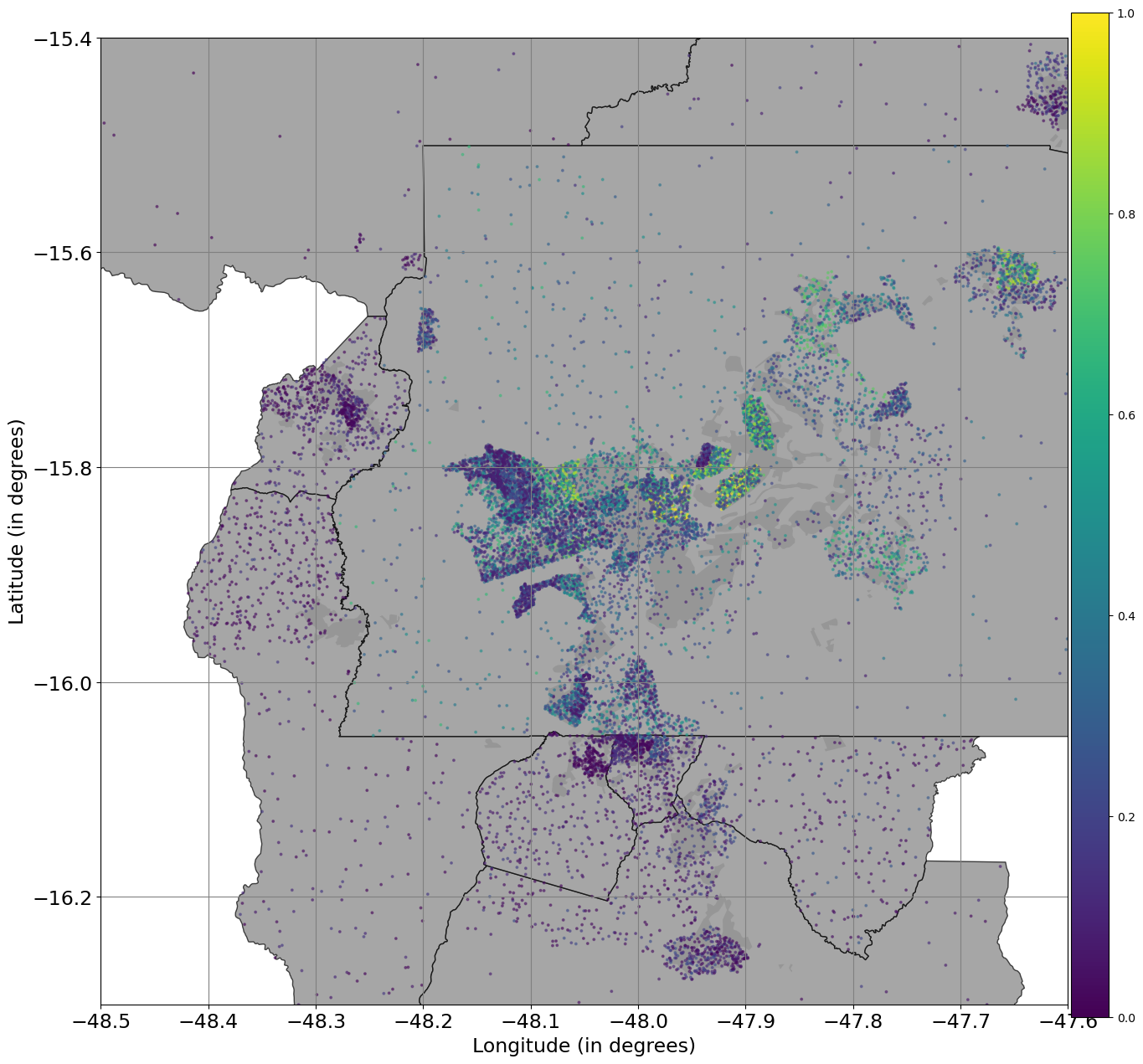}}
\caption{Comparison between house prices for real empirical data and simulated data for the case of Brasília, Brazil. Real data is drawn from public internet offers (2020). Simulated data follows default run parameters and refers to the last month of the simulation (2020) and is an illustration of a single run. Empirical data shows higher prices in the planned central area of Brasília, plus portions of more expensive places in the immediate of the wings (Sudoeste and Noroeste neighborhoods). A cluster of intermediate priced houses can be seen towards the west (Águas Claras), and cheaper further west. Simulated data shows house prices are also higher in the wings of the airplane and at the high-priced neighborhoods. The western portion is more mixed in the simulation, with lowest prices in the extreme west of the main urban area. Neighboring municipalities, excluded from the central Federal District, have cheaper houses.}
\label{fig:comparison}
\end{figure}

\begin{figure}[!t]
\centering
\includegraphics[width=.7\textwidth]{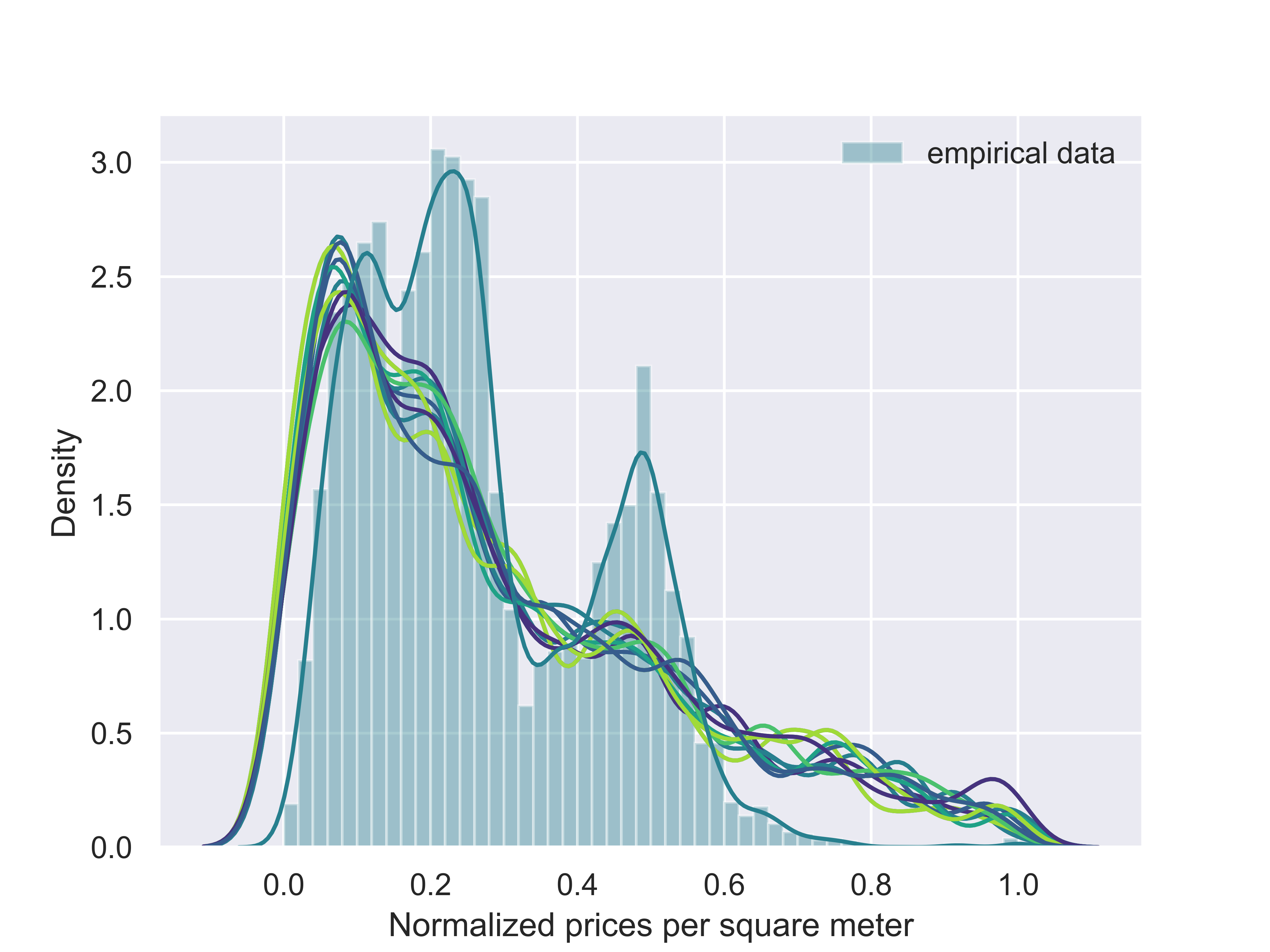}
\caption{Histogram of empirical house prices and simulated data (20 runs). The distribution of empirical data prices (filled histogram bars) is heterogeneous with a concentration of prices around the 10th and 30th percentile and another one around the median. The simulation data (colored single lines) is more homogeneous and has a higher number of observations at the higher parts of the distribution, thus showing more expensive houses comparatively to the empirical data.}
\label{fig:hist_comparison}
\end{figure}

We compare normalized prices over space for Brasília using data from the last month of the simulation (see figure \ref{fig:comparison}). The comparison aims at providing evidences that the generating mechanisms of relative prices across the metropolitan region are compatible with observed data. Results are spatially similar in the sense that the simulation is able to replicate a more expensive area in the central, planned area of Brasília, and lower priced areas in the neighboring municipalities and in the western portion of Brasília. Second-tier cluster of prices, especially at the northeastern area are more expensive in the simulation when compared to empirical data. That leads to a more homogeneous distribution in the simulated data compared to a more clustered, double-peaked distribution for the empirical data (see Figure \ref{fig:hist_comparison}). Observed prices are more volatile and ragged comparatively to simulated prices which are more continuous with less pronounced peaks. Simulated data also follows more closely the location of firms. 

However, considering that the description of properties size, quality or price is not included in the model, the similarity we were able to achieve, given a market that includes heterogeneous properties (central, small one-bedroom high valued properties, but also, large, distant, sophisticated properties) seem to be sufficient to hold the comparison and thus serve the purpose of the model.

Finally, the general centralized, richer configuration of the metropolitan region with higher unemployment, lower wages, and poorly more homogeneous peripheries is also reproduced in the simulation (see Table \ref{tab:parameters}).

\section{Results}\label{results}

\begin{figure}[!t]
\centering
\subfloat[GDP]{\includegraphics[width=.47\linewidth]{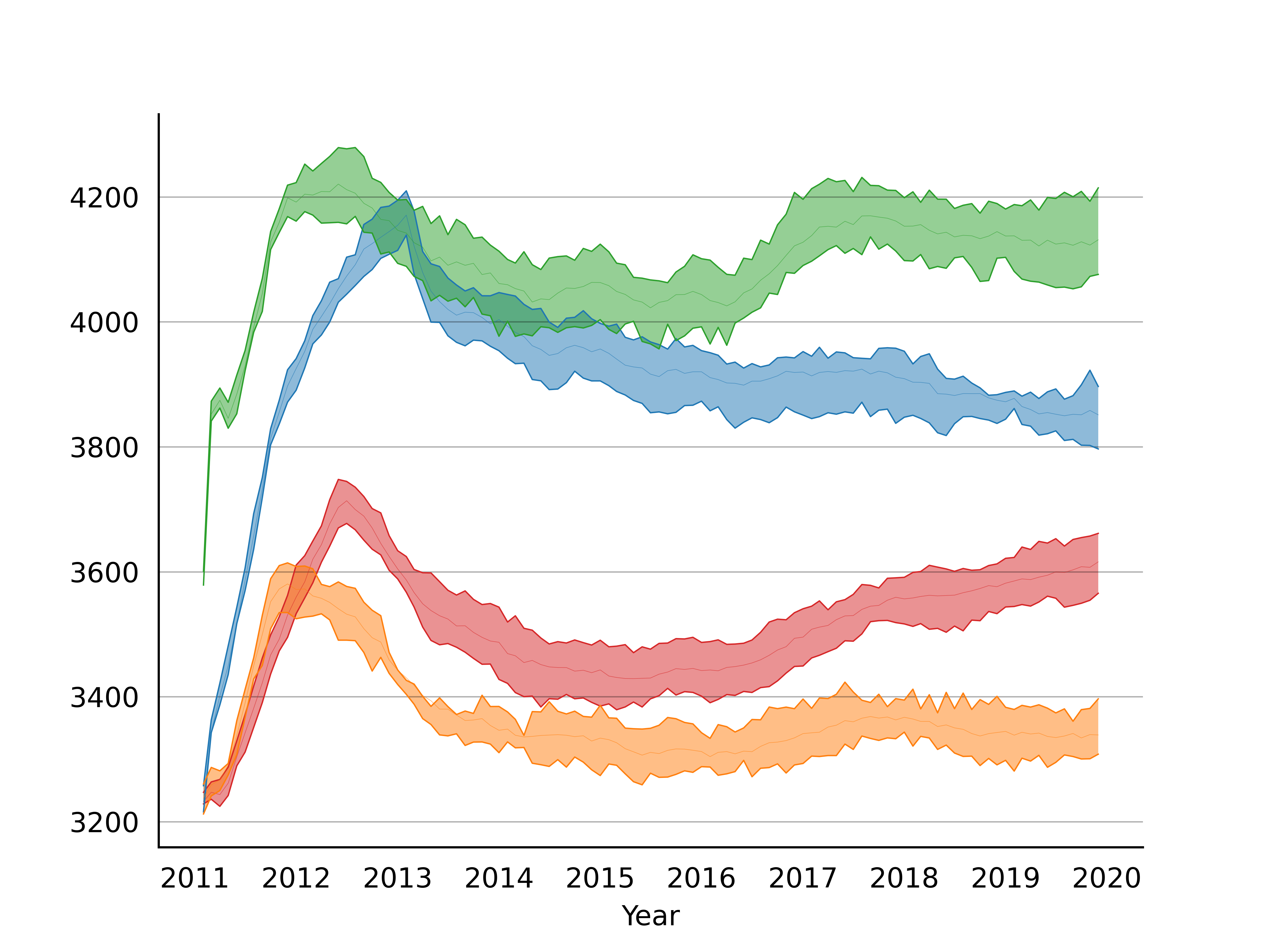}}\qquad
\subfloat[Gini coefficient]{\includegraphics[width=.47\linewidth]{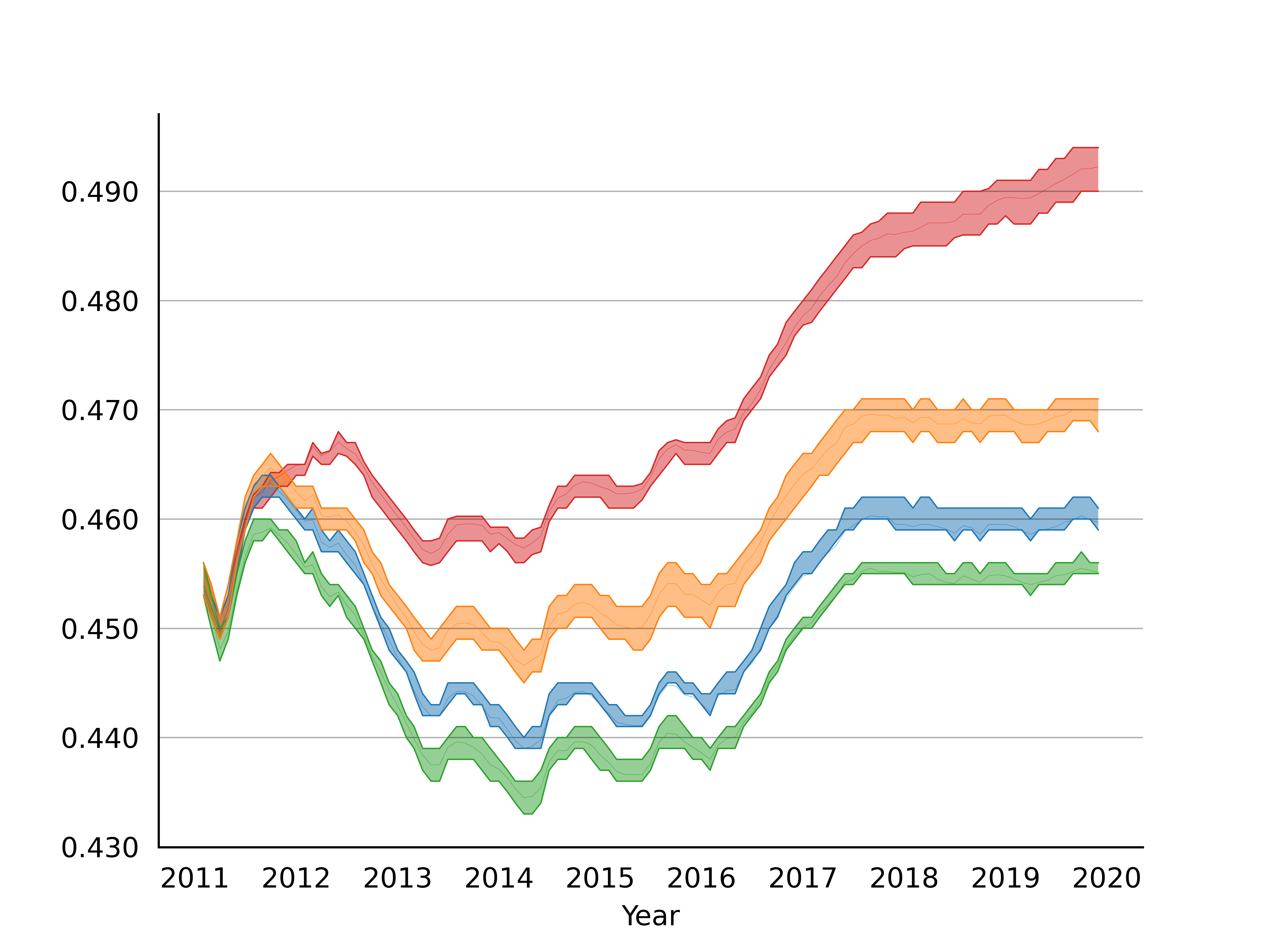}}\qquad
\subfloat[Quality of Life Index]{\includegraphics[width=.47\linewidth]{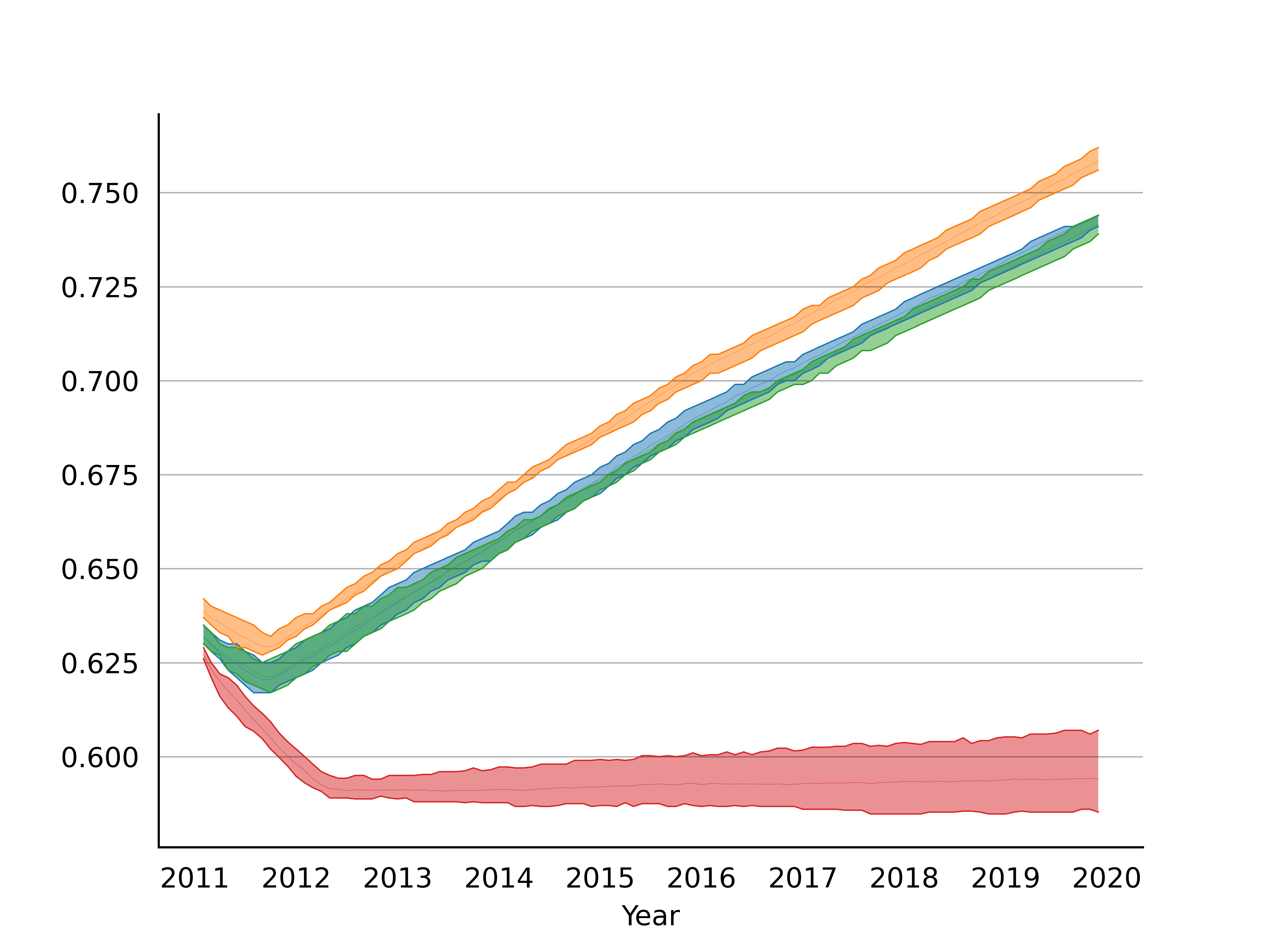}}\qquad
\subfloat[House prices]{\includegraphics[width=.47\linewidth]{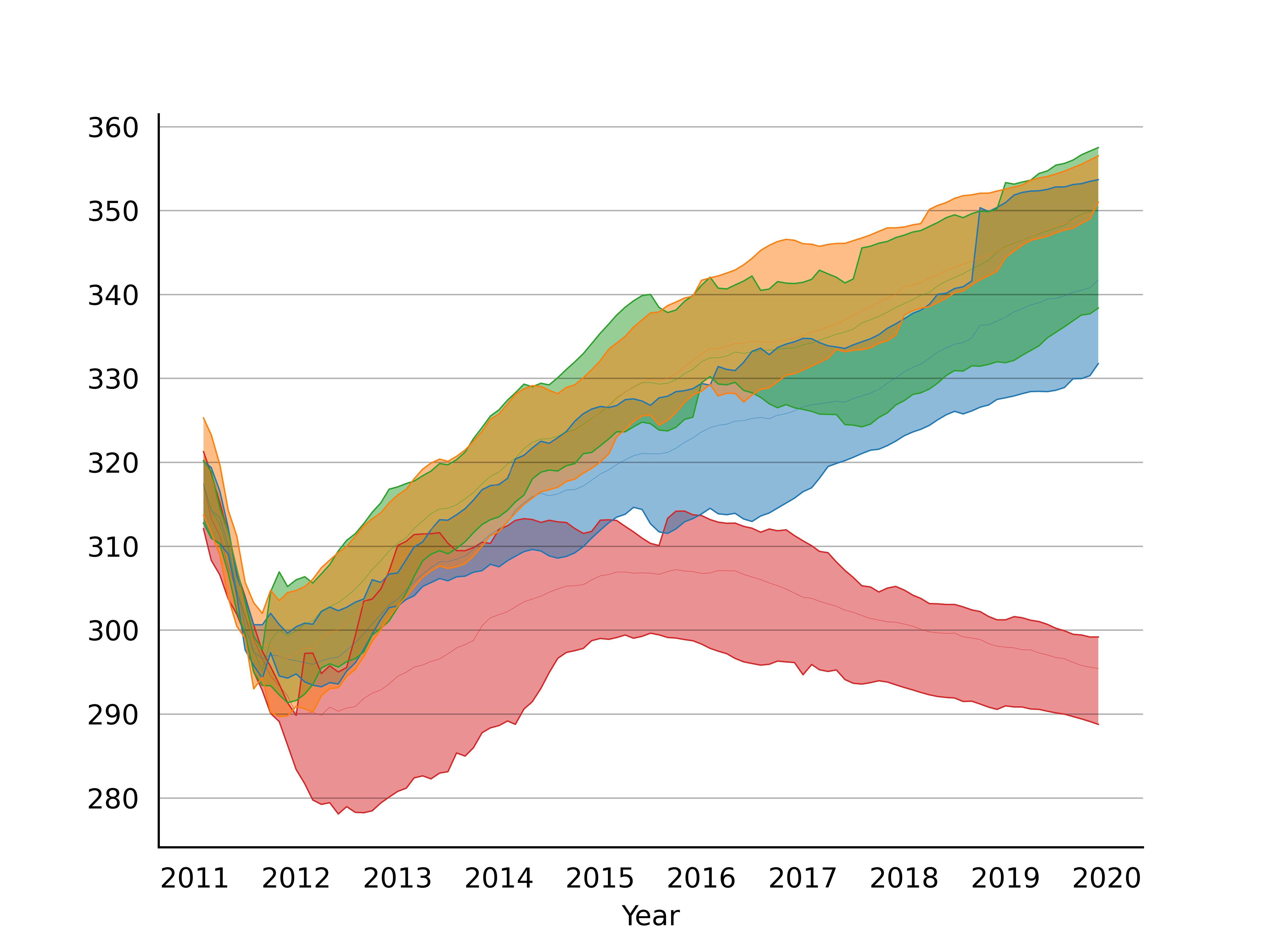}}\qquad
\subfloat[Percentage households defaulting on rent]{\includegraphics[width=.47\linewidth]{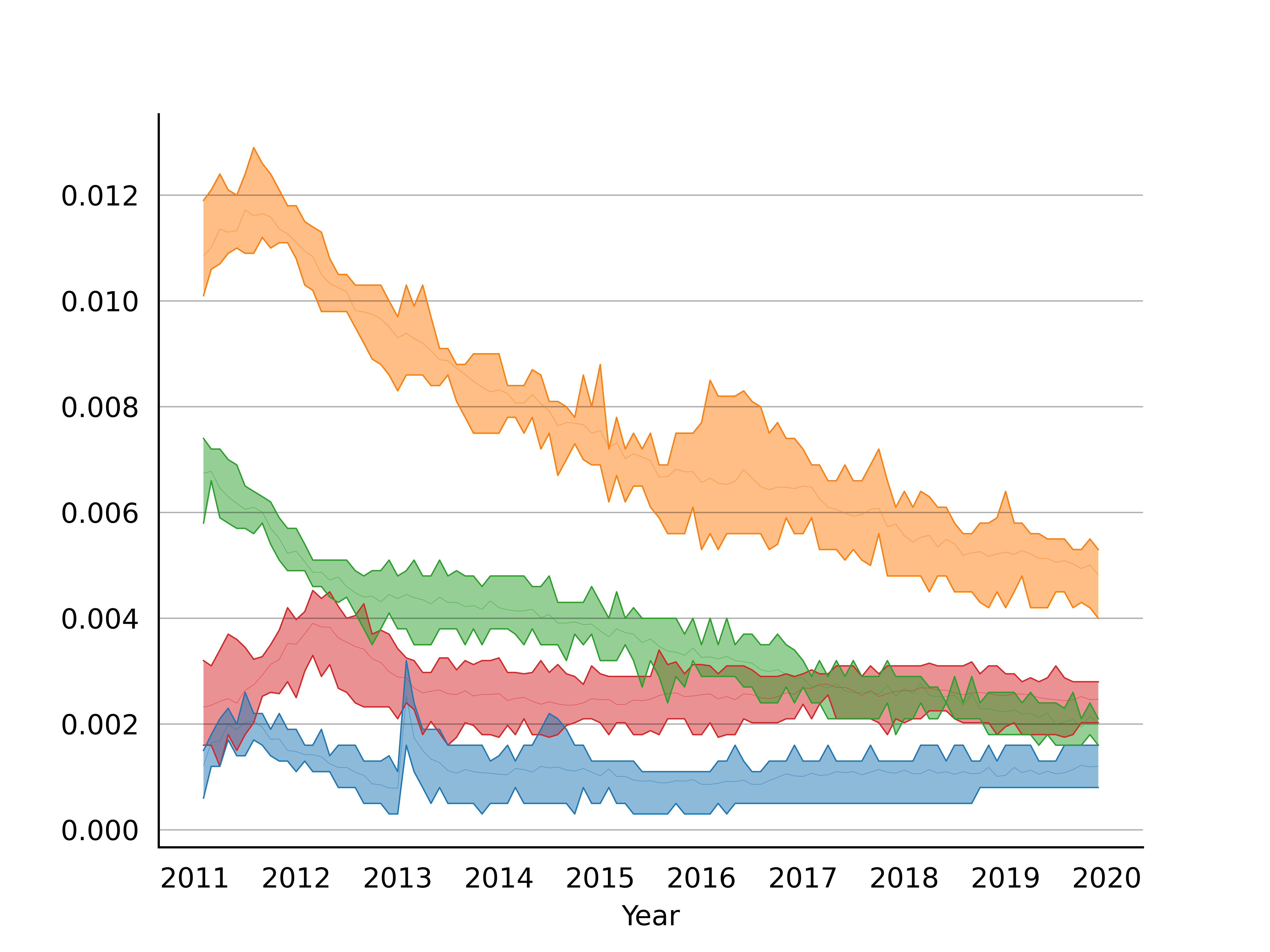}}\qquad
\subfloat[Percentage households with no consumption]{\includegraphics[width=.47\linewidth]{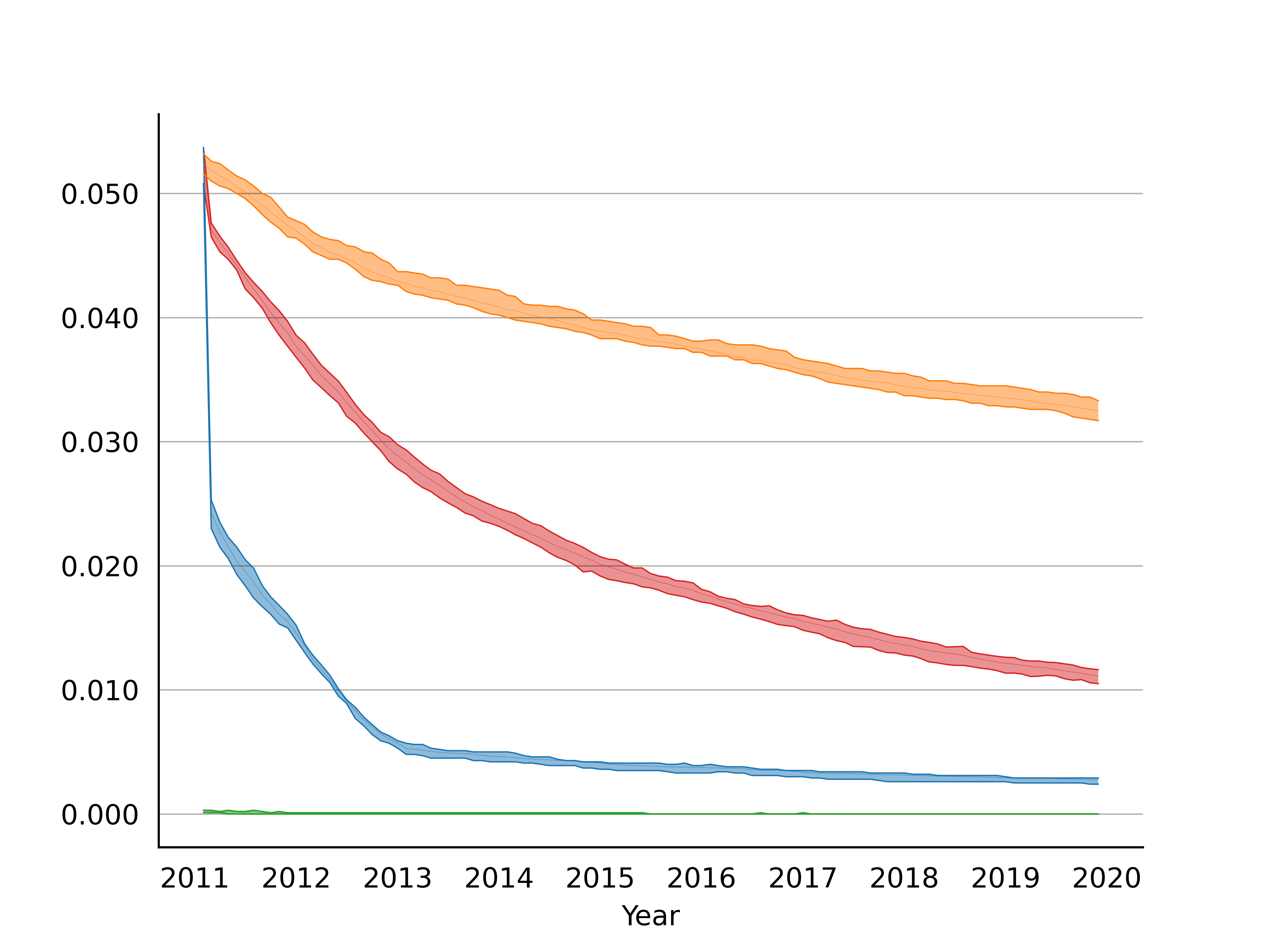}}
\caption{Results of policy experiment. \textbf{Red} refers to the Property Acquisition policy, \textbf{Blue} to Rental Vouchers, \textbf{Green} to Monetary aid and \textbf{Orange} is the No-Policy baseline. Monetary Aid is the policy application with a better performance both at the GDP level, but also with a reduced Gini coefficient indicator. Property Acquisition, however, seems to deteriorate inequality and contributes to GDP increase in a smaller proportion than the other policy tests, even worsening households' permanent income levels. Rental vouchers -- for obvious reasons -- result in the lowest number of households defaulting on rent, whereas Monetary aid contributes primarily to the maintenance of goods consumption, thus enforcing that households do not go a month without consuming.}
\label{fig:results}
\end{figure}

Results clearly show that, from an endogenous perspective, the best policy seems to be the Monetary aid, e.g., the distribution of a lower amount of aid directly to a larger number of households (see figure \ref{fig:results}). 

Given the endogenous amount of funding distributed towards the policies, it is up to the internal processes of the model how to foster the economy forward. When the Monetary aid policy is introduced, more households are able to consume larger quantities at the goods and services market. The resulting sales and revenues allow for firms to pay higher wages. This leads to an expected slight increase in prices. Even then, a smaller number of households fail to pay their rents or their mortgages or go a month without consuming goods and services. Overall, that leads to a much lower inequality with a higher GDP achievement.

The Rental voucher policy is a somewhat intermediate alternative with a much smaller number of households being supported. On average, Monetary aid policy helps 2,358.05 households monthly per municipality, compared to 47.5 for the Rental vouchers and 12.37 for the Property acquisition program.\footnote{That considers the standard run parameter of 1\% of the population of the metropolitan region and the policy coefficient ($\delta$) of .2 over the available funding for each municipality to apply on the policy. Alternative values for $\delta$ did not alter the results.} Even then, Rental policy seems to achieve lower inequality which is comparable to that of the Monetary aid, although with not as much an increase in GDP. 

Property acquisition seems to be the policy that performs worst. It lowers overall mean household permanent income compared to the other policies and to the No-policy baseline. Further, it sharply increases inequality -- as it provides an asset to a small number of households. On the positive side, it seems to increase firms (construction firms) total assets in a more pronounced way than the other two policy alternatives as well as it helps decrease house prices.  

The No-policy baseline results are provided as a comparison of the performance of the model. Consider that whereas the funding separated for the policy programs is reinvested in absolute terms, the investment when there is No-policy is made in full via the linear transformation of the $\psi$ parameter. Conversely, among the policies' alternatives, the exact same processes, procedures, and parameters run each policy scenario. That makes the policy alternatives highly comparable among themselves. 

\begin{figure}[!t]
\centering
\subfloat[Property acquisition policy]{\includegraphics[width=.47\textwidth]{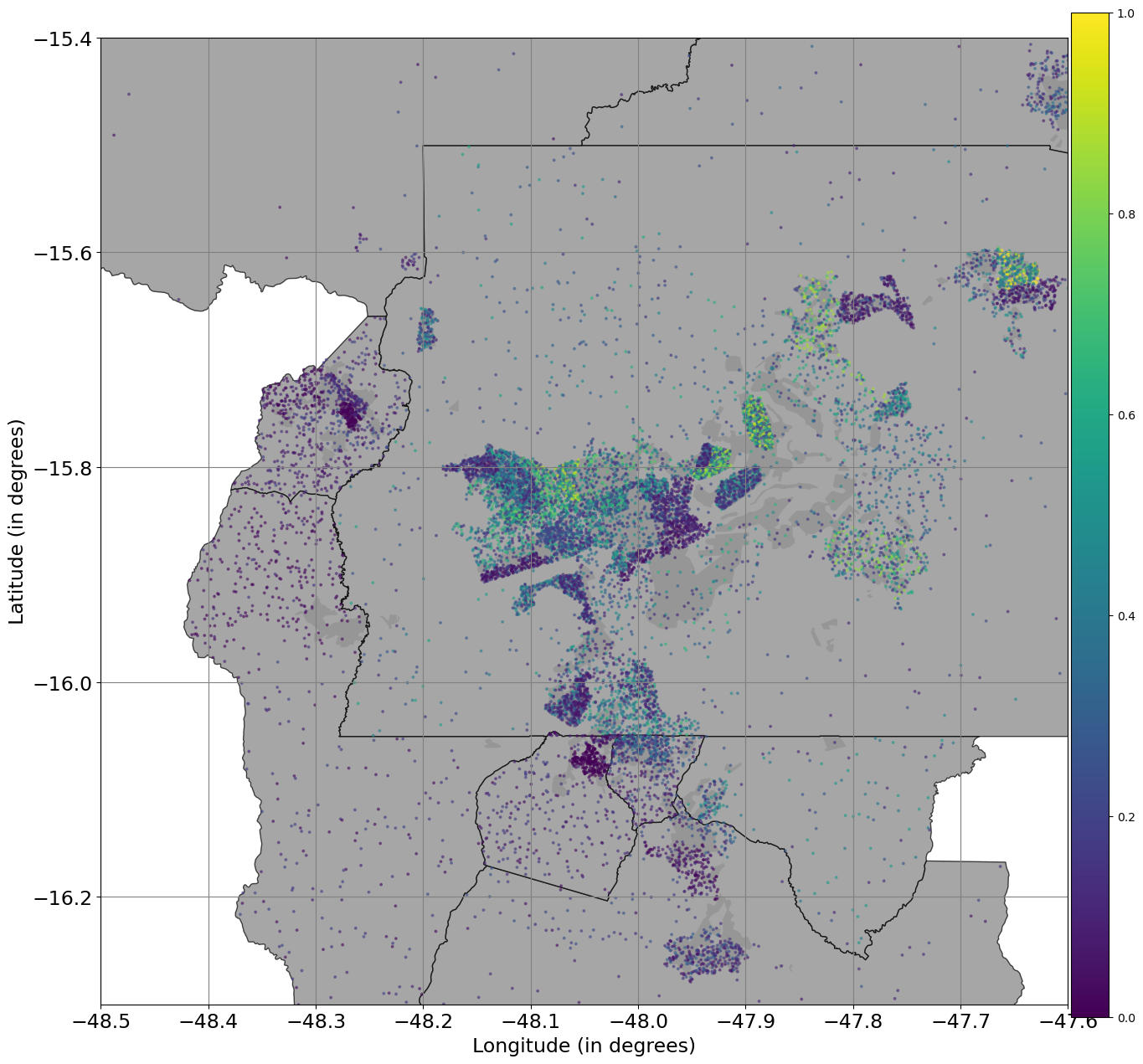}}\qquad
\subfloat[Rent Voucher policy]{\includegraphics[width=.47\textwidth]{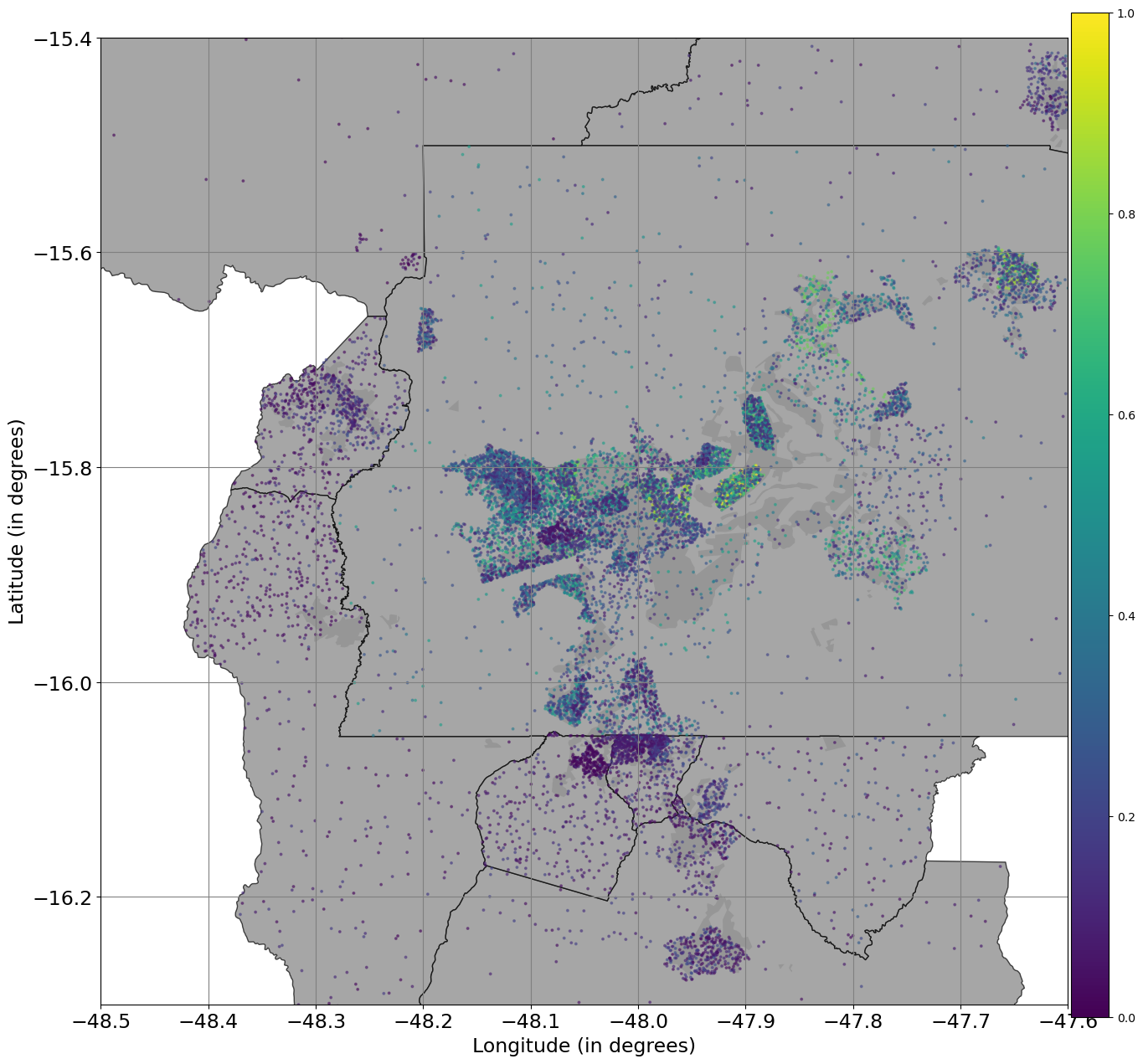}}\qquad
\subfloat[Monetary aid policy]{\includegraphics[width=.47\textwidth]{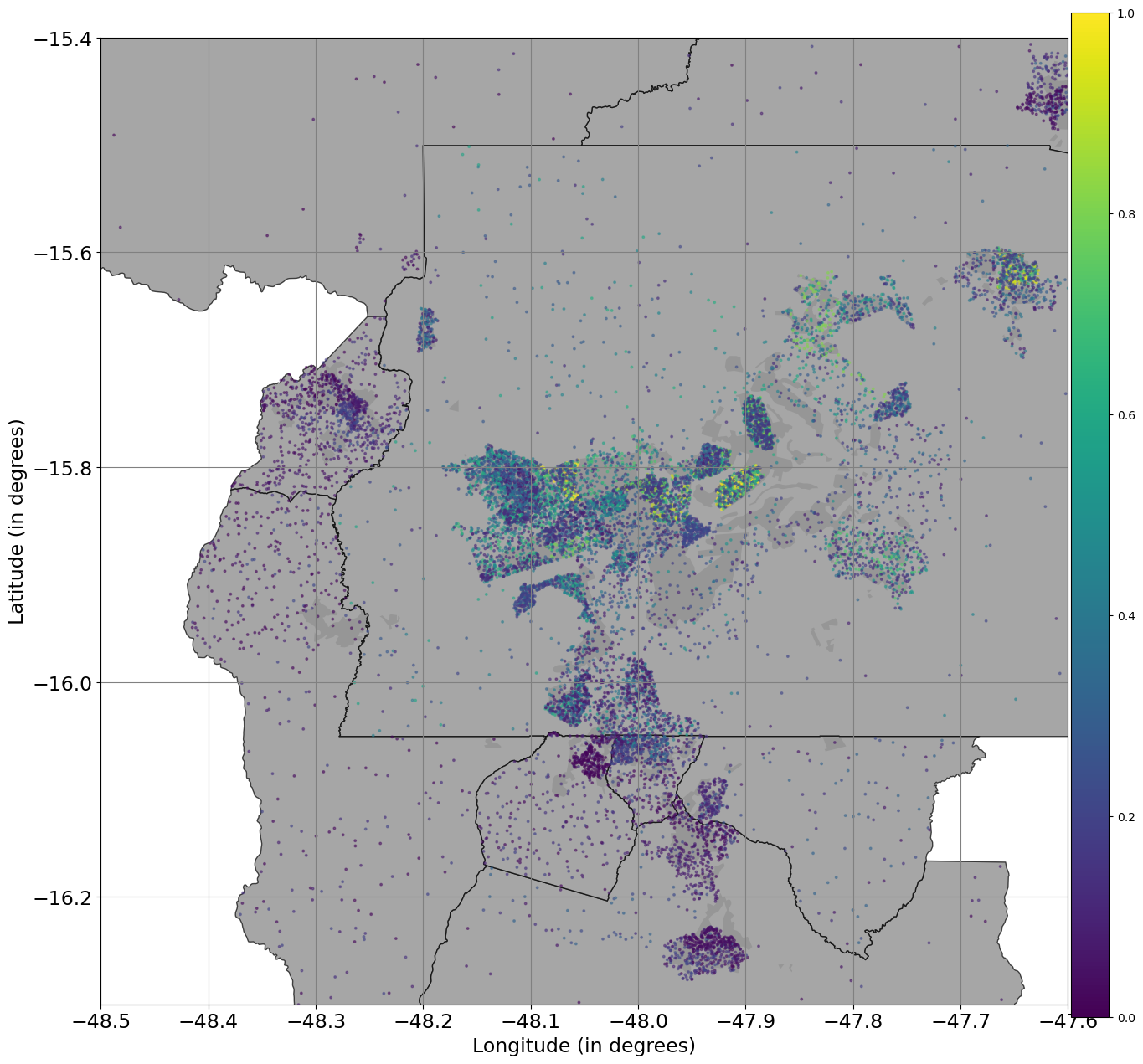}}
\caption{Illustration of spatial results for house prices in alternative policy scenarios for the case of Brasília, Brazil. Single run (2010-2020), prices at the end of the period. Although there are some spatial differences among the simulated  house prices across the policies, Monetary aid policy seem to have a lower number of cheaper (dark violet) priced houses, which reflects in the Gini results (Table \ref{fig:results}) }
\label{fig:spatial_results}
\end{figure}

Considering a spatial analysis of house prices over the policy alternatives (see figure \ref{fig:spatial_results}), Property acquisition and Monetary aid seem to result in a slightly more disperse and homogeneous distribution of prices, comparatively to Rent voucher and No-policy baseline, especially when the western populous region of Brasília and its southern border are considered. These results are confirmed by the analysis of the Gini coefficient for each municipality for the baseline case versus the application of the policy experiments (see table \ref{tab:gini_municipalities}).

\begin{table}[!t]
\centering
\begin{tabular}{lrrrr}
\toprule
Municipalities -- Brasília RM & Property & Monetary        & No-policy & Rental          \\
\midrule
Águas Lindas de Goiás & 0.4138   & \textbf{0.3538} & 0.3753    & 0.3628          \\
Cidade Ocidental      & 0.4238   & \textbf{0.3863} & 0.3961    & 0.3888          \\
Formosa               & 0.4317   & \textbf{0.3850} & 0.3996    & 0.3911          \\
Luziânia              & 0.4484   & \textbf{0.3961} & 0.4069    & 0.4002          \\
Novo Gama             & 0.4427   & \textbf{0.3761} & 0.3994    & 0.3826          \\
Padre Bernardo        & 0.3905   & \textbf{0.3414} & 0.3694    & 0.3559          \\
Planaltina            & 0.4451   & \textbf{0.4105} & 0.4303    & 0.4235          \\
Santo Antônio do Descoberto  & 0.3953   & 0.3436          & 0.3597    & \textbf{0.3430} \\
Valparaíso de Goiás         & 0.4444   & \textbf{0.3983} & 0.4179    & 0.4051          \\
Brasília                    & 0.4854   & \textbf{0.4467} & 0.4619    & 0.4521\\          
\bottomrule
\end{tabular}
\caption{Simulation data of Gini coefficient for the municipalities of Brasília Metropolitan Region. The indicator for the No-policy baseline case reflects the highest inequality in Brasília, followed by the more populous municipalities of Planaltina, Valparaíso de Goiás, and Luziânia, as observed in empirical data. Simulation results for Monetary aid and Rental vouchers show reduced intramunicipal inequality whereas Property acquisition generates the opposite effect.}
\label{tab:gini_municipalities}
\end{table}

Policy scenarios were also run for longer periods 2010--2030, keeping some exogenous data such as mortgage rates constant. Additionally, a very long run, based on census 2000 data was run for the period 2000-2030, using the same parameters that were calibrated for the 2010--2020 default configuration. In both cases, Gini coefficient maintains a trajectory of increasing inequality for the Property acquisition policy implementation whereas an unchanging path for the other alternatives. GDP, however, trends upwards also for Property acquisition, whereas a declining path is observed in the other scenarios. 

The robustness of the results are confirmed when we test the policy alternatives for four other middle-sized metropolitan regions of Brazil, from different regions, and with unique spatial, household, and qualification characteristics. Averaging results of six indicators for the full period and 20 simulation runs for each policy test are presented in table \ref{tab:comparison_among_cities}. Gini coefficient is lower on average for Monetary aid policy when compared to No-policy scenario for all cities, at the same time that GDP is higher. Household consumption is also higher for all cities. All policies applied generate higher inflation over the 10-year period, except for the case of Porto Alegre. Unemployment does not seem to be affected by policy, with results remaining stable for all cities across all policies. Finally, house prices seem to be lower for policy implementations, with more pronounced results in the case of Property acquisition, except for Porto Alegre that observes the lowest price on average for the Rent voucher case. 

\begin{table}[!t]
\centering
\begin{tabular}{lrrrr}
\toprule
              & \multicolumn{4}{c}{\textbf{Gini Coefficient}}    \\
\midrule
              & Property        & Rent voucher    & Monetary aid    & No-policy \\
Brasília       & 0.47            & \textbf{0.45}   & \textbf{0.45}   & 0.46 \\
Belo Horizonte & 0.42            & 0.41            & \textbf{0.4}    & 0.41 \\
Campinas       & 0.44            & 0.42            & \textbf{0.41}   & 0.42 \\
Fortaleza      & 0.44            & \textbf{0.42}   & \textbf{0.42}   & 0.43 \\
Porto Alegre   & 0.44            & 0.43            & \textbf{0.42}   & 0.43 \\
\midrule
              & \multicolumn{4}{c}{\textbf{GDP}}    \\
              \midrule
              & Property        & Rent voucher    & Monetary aid    & No-policy       \\
Brasília       & 3410.3          & 3768.7          & \textbf{3814.3} & 3298.3 \\
Belo Horizonte & 5664.1          & 6174.4          & \textbf{6326.6} & 5520.1 \\
Campinas       & 3545.7          & 3782.7          & \textbf{3905.1} & 3312.4 \\
Fortaleza      & 3684.9          & 3954.9          & \textbf{4025.4} & 3545.3 \\
Porto Alegre   & 4052.6          & 4258.0          & \textbf{4341.3} & 3882.2  \\
\midrule
              & \multicolumn{4}{c}{\textbf{Household consumption}}            \\
              \midrule
              & Property        & Rent voucher    & Monetary aid    & No-policy       \\
Brasília       & 0.25            & \textbf{0.31}   & \textbf{0.31}   & 0.25  \\
Belo Horizonte & 0.28            & 0.34            & \textbf{0.35}   & 0.28  \\
Campinas       & 0.32            & 0.38            & \textbf{0.4}    & 0.31   \\
Fortaleza      & 0.29            & 0.34            & \textbf{0.35}   & 0.29    \\
Porto Alegre   & 0.31            & 0.36            & \textbf{0.37}   & 0.31      \\
\midrule
              & \multicolumn{4}{c}{\textbf{Price index}}       \\
              \midrule
              & Property        & Rent voucher    & Monetary aid    & No-policy       \\
Brasília       & 1.39            & 1.49            & 1.48            & \textbf{1.38}   \\
Belo Horizonte & 1.74            & 1.96            & 2.01            & \textbf{1.71}   \\
Campinas       & 1.56            & 1.76            & 1.83            & \textbf{1.55}   \\
Fortaleza      & 1.41            & 1.54            & 1.57            & \textbf{1.39}   \\
Porto Alegre   & \textbf{1.57}   & 1.81            & 1.87            & 1.58            \\
\midrule
              & \multicolumn{4}{c}{\textbf{Unemployment}}        \\
              \midrule
              & Property        & Rent voucher    & Monetary aid    & No-policy   \\
Brasília       & 0.11            & 0.11            & 0.11            & 0.11    \\
Belo Horizonte & 0.13            & 0.13            & 0.13            & 0.13  \\
Campinas       & 0.06            & 0.06            & 0.06            & 0.06   \\
Fortaleza      & 0.1             & 0.1             & 0.1             & 0.1      \\
Porto Alegre   & 0.08            & 0.08            & 0.08            & 0.08    \\
\midrule
              & \multicolumn{4}{c}{\textbf{House prices}}      \\
              \midrule
              & Property        & Rent voucher    & Monetary aid    & No-policy       \\
Brasília       & \textit{308.56} & 322.86          & 320.71          & \textbf{327.56} \\
Belo Horizonte & 223.44          & \textbf{225.13} & \textit{222.86} & \textbf{225.78} \\
Campinas       & 255.45          & 254.76          & \textit{249.55} & \textbf{260.09} \\
Fortaleza      & \textit{256.47} & 267.39          & 266.32          & \textbf{270.93} \\
Porto Alegre   & 261.24          & \textit{261.14} & 262.09          & \textbf{266.59} \\
\bottomrule
\end{tabular}
\caption{Results for six indicators, averaged over the 2010--2020 period and 20 simulation runs for each city and each policy alternative. Although the cities are very diverse and different among themselves, the results of the simulations seem to maintain the same trend of results.}
\label{tab:comparison_among_cities}
\end{table}

\subsection{Intuition, discussion and future work}

One of the essential differences among the policies tested is that Monetary aid is provided for a larger number of households relative to Property acquisition and Rental voucher. Rental voucher performs close to Monetary aid and further from Property acquisition for different indicators. In practice, that would mean that a social welfare policy program, such as the Bolsa Família, instated in 2003, and expanded in subsequent administrations, may be more beneficial than the housing program MCMV. Specifically to provide affordable housing, rental mechanisms may be a better policy when compared to property distributions. 

We believe that a mechanism that plays a relevant role in the simulation is that when buying houses and handing them over to households immobilizes capital whilst supporting construction firms. This capital then returns back to the system as savings (which may be spent in goods), in the increase of housing stock and in construction firms' workers' salaries. However, that feedback might be slower in comparison to Monetary aid and Rental voucher policies. 

Further, this capital immobilization impacts collection of taxes. The dynamics of Monetary aid and Rental vouchers enable the maintenance of the tax collection trends. Property acquisition, however, does not seem to maintain the level investments on municipalities, thus affecting the updates on the Quality of Life Index. This, in turn, also affects house prices.

The analysis of alternative policy scenarios is made considering the perspective of municipalities and local policy-makers. The question often posed by politicians is: "given an amount of financial resources, which policy will mostly benefit citizens?" Along with this question, we believe that PS2 contributes with the ability to observe endogenous effects across different aspects of social life. As such, policies across different areas (housing or social benefits) may be evaluated relative to one another. 

Providing housing alone may not be sufficient if the benefit does not include jobs and services and access to the city, as shown by the policy program implemented through 2009--2019. Moreover, houses' characteristics make it an expensive good \citep{arnott_economic_1987} in a thin, volatile \citep{glaeser_extrapolative_2017}, complex market. 

All considered, if the municipality is seeking a specific housing policy, the simulations done suggest that a Rental voucher may be more beneficial. At the same time that provides families with house security, the policy additionally grants households with mobility \citep{8631423}, whilst keeping municipal investments relatively lower. However,  if the municipality is not seeking a housing policy \textit{per se}, but otherwise a general policy to invest its financial resources, a Monetary aid alternative might be comparatively better. 

PS2 is already a complex model. Nevertheless, we are considering a number of future developments. Firstly, the model can always be improved, and interactions more detailed. For PS2 specifically that might involve the credit market, the firms, the transport and education systems. Secondly, we also intend to publish further material that would include results for all 46 metropolitan regions, along with policy comparisons for all of them. Such publication would also include more details of the comprehensive sensitivity analysis we have performed. Finally, in terms of policy analysis, we would like to test whether simultaneous policies at different amounts of percentage of the available budget would make an optimal decision mix. Policies could also be financed exogenously, for example, by the Federal government. 
 
\section{Final considerations}

We apply an empirical, economic, spatial, open-source agent-based model that uses official data to generate households, firms, and municipalities that interact in the markets of labor, credit, real estate and goods, and services to endogenously test alternative, cross-sectors policy programs. Poor households -- those from the lowest fifth of the distribution -- are registered at their municipalities and organized according to their observed income levels. Municipalities collect funds via taxes on consumption, labor, firm's profits, investments, property transactions, and tax on properties. Regularly municipalities invest in a general improvement of quality of life. When applying policies, the municipal body alternately reserves a fifth of its funds to (a) acquire and distribute properties for poorest households, (b) provide 24-month rental vouchers or (c) simply divide the monthly available resources with all of the registered households. Results hold for a number of different parameter intervals, rules and applied processes, and metropolitan regions tested. Within the context of the model, the monetary aid performs better than the alternatives in nearly all of the indicators used. Mainly, monetary aid reduced inequality at the same time that it increased overall economic output.

PS2 benefits from a  number of previous modeling works on macroeconomics, housing markets, transport, and land-use change. We tried to incorporate most of the benchmarks and procedures both in modeling and in communicating. The confidence in PS2 comes from the sustainable robustness it has shown on top of added mechanisms, data, parameters, and markets. The confidence in the results come from the theoretical comparison among the alternative choices. 

The aim of the paper is not, however, to recommend an extinction of housing acquisition and distribution to poorest families. We believe PS2 only demonstrates that a given policy -- however focused it might be -- might endogenously contribute more to the economy when dispersing more funds, rather than concentrating them in fewer households. 

Nevertheless, when facing a specific housing program with strict lack of shelter, rental vouchers might benefit a larger number of people and result in greater gains for the society as a whole.\footnote{We also noticed that poor families recurrently -- although slightly diminishing in numbers -- accessed the Rental program. As such, other structural policies might be needed.}  

Finally, we believe that the contributions of PS2 surpasses the policy analysis itself. We provide a comprehensive empirical, documented, and robust agent-based model that is open source and modular and that might facilitate further research questions. The difficulties in working with PS2 and other ABMs is probably due to its flexibility and adaptation. We believe PS2 is nearly ready to be expanded and used in a series of other analyses, for instance: (a) urban mobility -- given location of workers and firms, along with endogenous characteristics of both; (b) social mobility -- given the dependence of workers productivity on qualification and its current static nature; (c) migration and newly formed households; (d) inheritance -- for analyses with longer periods of time; (e) a more detailed credit mechanism and authority ruling; (f) urban zoning and regulation; (g) amenities, neighborhood perceptions and its influences on prices. Given proper data and initial agent generation, PS2 might also serve the purpose of fostering understanding of long-term market prices and behaviors. 

\section{Model Documentation}

\begin{enumerate}
    \item PS2 runs in \texttt{python v.3.8.5} and requires spatial libraries, such as \texttt{shapely, gdal, descartes, fiona, and geopandas=0.7.0}, and numerical and fast-processing ones, such as \texttt{numba, joblib and scipy}.
    \item Fork and clone the repository from \href{https://www.comses.net/codebase-release/bc5b116a-5fdf-4b6f-837a-a7978ab34268/}{ComSES Open ABM: https://www.comses.net/codebase-release/bc5b116a-5fdf-4b6f-837a-a7978ab34268/}.
    \item After installation of requirements, make sure to alter \texttt{conf/run.py} and adjust \texttt{OUTPUT\_PATH} for your saving folder. 
    \item You may use \texttt{--runs} and  \texttt{--cpus} to specify number of runs and CPUs to be used in a simulation. Default uses all cores in your computer. 
    \item A simple run with two CPUs and for 10 simulation runs is run with the command: \texttt{python main.py -c 2 -n 10 run}
    \item Sensitivity analysis on parameters may be run with \texttt{python main.py sensitivity ALPHA:0:1:7}, for example. It may also include options -c and -n. The syntax is \\\texttt{sensitivity PARAM:INITIAL\_VALUE:LAST\_VALUE:NUMBER\_INTERVALS}
    \item To test for the Policies, you may run: \texttt{python main.py sensitivity POLICIES}
    \item Plots are generated automatically. Some options for running, saving and plotting are available in the \texttt{conf/run} folder.
\end{enumerate}

\bibliographystyle{IEEEtranN}
\bibliography{PS2} 
\section{Appendix}
\label{appendix}

\subsection{Input data}
See table \ref{tab:input_data} for the parameter values used on the standard simulation run.

\begin{table}[!t]
\centering
\resizebox{\textwidth}{!}{%
\begin{tabular}{p{.2\textwidth}p{.13\textwidth}cp{.13\textwidth}ccccp{.2\textwidth}}
\toprule
Input Data for Metropolitan Region of Brasília &
  Source &
  Period &
  Observations * cols &
  Max &
  Min &
  Mean &
  Standard-deviation &
  Observations \\
 \midrule
Urban proportion &
  IBGE sidra:table 202 &
  2010 &
  10 &
  1 &
  0.3898 &
  0.8824 &
  0.1848 &
  Brasília + 9 mun. \\
Females per mun. per age &
  IBGE sidra: table 1378 &
  2010 &
  10 (mun.) * 101 (year of age) &
  30467 &
  0 &
  1772.6 &
  2964.39 &
  Ages from 0 to 100. Statistics for all ages. \\
Males per mun. per age &
  IBGE sidra: table 1378 &
  2010 &
  10 (mun.) * 101 (year of age) &
  27115 &
  0 &
  1653.78 &
  2846.47 &
  Age from 0 to 100. Statistics for all ages. \\
People per region per gender per age &
  IBGE: census blocks &
  2010 &
  17978 &
  1367 &
  0 &
  192.47 &
  194.08 &
   \\
Number of members per family per region &
  IBGE: census blocks &
  2010 &
  89 &
  3.89 &
  2.42 &
  3.41 &
  0.26 &
   \\
Population estimates &
  IBGE &
  2001-2019 &
  10 * 17 &
  3039444 &
  22108 &
  353497.94 &
  99428.83 &
  Statistics for all years. \\
Geographic shapefiles &
  IBGE: geoftp &
  2010 &
  89 &
   &
   &
   &
   &
  Intraurban regions. Brasília: 51, neighboring mun: 38\\
Number firms per region &
  Ministry of Labor: RAIS &
  2012 &
  89 &
  7446 &
  3 &
  660.84 &
  1106.12 &
   \\
Number firms per region &
  Ministry of Labor: RAIS &
  2017 &
  89 &
  8299 &
  3 &
  750.27 &
  1158.48 &
   \\
Cumulative prob. highest level instruction per region &
  IBGE: census blocks &
  2010 &
  89 &
  1 &
  0.1814 &
  0.8177 &
  0.0771 &
  5 qualification levels \\
Mortgage real interest &
  BACEN: series 25497 &
  2010-2020 &
  241 &
  0.016 &
  0.0001 &
  0.0076 &
  0.0034 &
   \\
Fertility per age per calendar year &
  IBGE &
  2010-2020 &
  35 (age years) * 21 (calendar years) * 2 states &
  0.0306 &
  0.00015 &
  0.0099 &
  0.00125 &
  State-level data age 15-49 \\
Mortality per gender per age per calendar year per state &
  IBGE &
  2010-2020 &
  2 (gender) *  111 (age years) * 21 (calendar years) * 2 states &
  1 &
  0.000105 &
  0.1011 &
  0.02406 &
  State-level data age 0-110 \\
FPM: municipal tax transfers &
  Ministry of Economics &
  2000-2016 &
  171 &
  135441938.2 &
  2206964.75 &
  22061509.85 &
  22369980.67 &
   \\
HDI-M &
  FJP and Ipea &
  2010 &
  10 &
  0.824 &
  0.651 &
  0.7087 &
  0.0516 &
  atlasbrasil.org.br/2013/\\
  \bottomrule
\end{tabular}%
}
\caption{Table of input data entered in the model. All data refers to demographic and locational input for individuals, households and firms and come from official data. The repository of the model contains all the necessary data to run the model. No data referring to quality, size or price of houses enter the model. \textbf{Obs.}: All mentions of "region" refer to intraurban regions and mun. are the municipalities. IBGE refers the Brazilian Statistics Bureau, and sidra is their statistics catalog system. RAIS is the Annual list of mandatory employees data filled by companies and compiled by the Ministry of Economics, previously Ministry of Labor. BACEN refers to the Central Bank (www.bcb.gov.br). FJP refers to Fundação João Pinheiro (fjp.mg.gov.br), a think-tank, as well as Ipea -- Institute for Applied Economic Research (www.ipea.gov.br). HDI-M is the Municipal Human Development Index. FPM referst to the Fund of Municipal Transfers.}
\label{tab:input_data}
\end{table}

\subsection{Parameters}
See table \ref{tab:parameters} for the parameter values used on the standard simulation run.

\begin{table}[!t]
\centering
\begin{tabular}{llll}
\hline\noalign{\smallskip}
Parameters & Code name & Standard values & Tested intervals  \\
\noalign{\smallskip}\hline\noalign{\smallskip}
$\text{pop}$ & percentage of population & .01 & [.005, .03] \\
$\alpha$ & productivity exponent & .6 & [0, 1] \\
$\beta$ & productivity magnitude divisor & 10 & [1, 36] \\
$\iota$ & labor market & .75 & [0, 1] \\
$\eta$ & percentage distance hiring & .3 & [0, 1] \\
$\phi$ & perc. entering real estate market & .0045 & [0, .05] \\
$\sigma$ & hiring sample size & 20 & [1, 100]\\
$\varsigma$ & size market & 5 & [1, 20]\\
$\rho_{+}$ & capped top value & 1.3 & [1, 1.5] \\
$\rho_{-}$ & capped low value & .7 & [.5, 1] \\
$\tau$ & neighborhood effect & 3 & [0, 5] \\
$\gamma$ & max offer discount -- lower bound & .6 & [.5, 1] \\
$\kappa$ & on market decay factor & -.01 & [0, -.05] \\
$\pi$ & markup & .15 & [0, .3] \\
$\psi$ & municipal efficiency index & .00007 & [.00001, .0001] \\
$\nu$ & max loan bank percentage & .7 & [0, 1] \\
$\chi$ & loan payment to permanent income & .5 & [0, 1] \\
$n$ & construction cash flow -- n. months & 24 & [1, 36] \\
$\upsilon$ & lot cost & .15 & [.01, .3] \\
$\zeta$ & sticky prices & .7 & [.1, .9] \\
$\delta$ & policy coefficient & .2 & [0, .3] \\
$\theta$ & policy quantile & .2 & [.1, .3] \\
\noalign{\smallskip}\hline
\end{tabular}
\caption{Parameters used on standard simulation run for the case of Brasília metropolitan region, 2000-2010, minimum of 5 runs each.}
\label{tab:parameters}  
\end{table}

\end{document}